%% file: paper.tex
\documentclass[sn-mathphys, Numbered]{sn-jnl}

\usepackage{bigfoot}

\usepackage{comment}
\usepackage{enumitem}

\usepackage{graphicx}

\usepackage{tabularx}
\usepackage{caption}

\usepackage{amsmath}
\usepackage{amssymb}

\usepackage[ruled, vlined, linesnumbered]{algorithm2e}
\usepackage{todonotes}
\setuptodonotes{inline}

\usepackage{dune}

\usepackage{pgfplots}
\pgfplotsset{compat=1.9,legend style={font=\small}}
\usepgfplotslibrary{groupplots,dateplot}
\usetikzlibrary{patterns,shapes.arrows,shapes.geometric,plotmarks}

\usepackage{multirow, makecell}
\usepackage{subfig}

\input{macros}

\renewcommand{\vec}[1]{\mathbf{\boldsymbol{#1}}}

\begin{document}


\title{A new \dune grid
       for scalable dynamic adaptivity
       based on the \pforest software library}


\author[1]  {\fnm{Carsten} \sur{Burstedde}}\email{burstedde@ins.uni-bonn.de}
\author[1,2]{\fnm{Mikhail} \sur{Kirilin}}\email{kirilin@ins.uni-bonn.de}
\author*[3] {\fnm{Robert}  \sur{Kl\"ofkorn}}\email{robertk@math.lu.se}



\affil[1]{\orgdiv{Institute f\"ur Numerische Simulation},
\orgname{Rheinische Friedrich-Wilhelms-Universit\"at Bonn}, \city{Bonn},
\postcode{53115}, \country{Germany}}
\affil[2]{\orgdiv{Institute of Mathematics},
\orgname{Technische Universit\"at Berlin}, \city{Berlin},
\postcode{10623}, \country{Germany}}
\affil[3]{\orgdiv{Center for Mathematical Sciences}, \orgname{Lund University}, \orgaddress{Box 117}, \city{Lund}, \postcode{22100}, \country{Sweden}}

\abstract{\input{abstract}}


\date{July 2025}

\maketitle


\input{introduction}

\input{techdesign}

\input{implementation}

\input{numericalexperiments}

\input{conclusions}



\bibliography{bibliography,
ccgo_new,
carsten
}


%
\appendix

\input{installation}

\end{document}

%% file: macros.tex
\usepackage{bbm}
\usepackage{xspace}

\colorlet{darkorange}{red!70!yellow!70!black}
\colorlet{midorange}{red!50!yellow!80!black}
\colorlet{brightblue}{blue!90!red!60!white}

\newcommand{\libsc}{\texttt{sc}\xspace}
\newcommand{\pforest}{\texttt{p4est}\xspace}
\newcommand{\pforthree}{\texttt{p4est3}\xspace}
\newcommand{\tetcode}{\texttt{t8code}\xspace}

\newcommand{\pforestbalance}{\texttt{p4est\_balance}\xspace}
\newcommand{\pforestexchange}{\texttt{ghost\_exchange}\xspace}
\newcommand{\LAND}{\textbf{\upshape and}\xspace}

\newcommand{\pforestobject}[1]{\texttt{\upshape p4est\_#1\_t}\xspace}
\newcommand{\pforestforest}{\texttt{\upshape p4est\_t}\xspace}
\newcommand{\pforestconnectivity}{\pforestobject{connectivity}}
\newcommand{\pforestquadrant}{\pforestobject{quadrant}}
\newcommand{\pforestghost}{\pforestobject{ghost}}

\newcommand{\eqnlab}[1]{\label{eqn:#1}}
\newcommand{\eqnref}[1]{\eqref{eqn:#1}}
\newcommand{\seclab}[1]{\label{sec:#1}}
\newcommand{\secref}[1]{Section~\ref{sec:#1}}

\newcommand{\floors}[1]{\left\lfloor #1 \right\rfloor}

\newcommand{\hgrid}{{\mathcal{H}}}
\newcommand{\grid}{{\mathcal{G}}}
\newcommand{\entity}{E}
\newcommand{\elem}{\entity}
\newcommand{\neig}{K}

\newcommand{\id}{\rm id}

\newcommand{\vect}[1]{\boldsymbol{#1}}

\newcommand{\vecu}{\vect{u}}

\newcommand{\ics}{\vect{x}}

\newcommand{\interset}{\mathcal{I}_{\elem}}

%% file: abstract.tex
In this work we extend the \dune solver library with
another grid interface to the open-source \pforest software.
While \dune already supports about a dozen different mesh implementations
through its mesh interface \dunegrid, we undertake this new coupling effort
in order to inherit \pforest's practically unlimited MPI scalability
as well as its relatively thin data
structures, and its native support for multi-block (forest) mesh topologies
in both 2D and 3D.

The presented implementation is compared to an existing implementation based on
\dunealugrid for a variety of challenging test examples in a parallel
environment. The numerical experiments show that the implementation presented
here is outperforming \dunealugrid in terms of scalability. In addition, an
alternative balancing strategy is presented to ensure 2:1 balancing across
element faces showing improved performance compared to the existing \pforest balance strategy
in the numerical examples considered in this work.

%% file: introduction.tex
\section{Introduction}
\seclab{introduction}

Modularity and the separation of concerns are cornerstones of any sizable
software development effort.
Since various software libraries and frameworks developed in the applied
mathematics community span hundreds of thousands or even millions of lines
of code, the issue of how to best interface to any of them remains current
and relevant.
One dividing line observed frequently arises between the discretization
layer and the mesh management layer of any PDE solver code.
Often, the mesh layer is an abstraction in itself and offers to interface to
various backends on an even lower level of the software stack.
Most details of the parallelization of the mesh layer can often be
encapsulated and hidden from higher levels of the code.

One relatively advanced but often necessary functionality with regard to
meshing is adaptivity, that is, the static (a-priori) and also the dynamic
(run-time) update of the mesh structure via local coarsening and refinement.
Implementing adaptivity in a parallel computation environment implies the
need for efficient load balancing of the mesh through repartitioning.
When building a numerical solver around such a functionality, there is
generally a two-way flow of information to and from the mesh layer:
The solver may inform the mesh layer of local criteria for adaptation and
weights for partitioning, while the mesh layer modifies and redistributes
the mesh structure itself in parallel, which must be related back to the
solver to update, recreate, renumber and/or transfer the numerical
variables.

The literature as well as the landscape of codes on the subject has become
rather extensive, so the following references can serve only as a tiny
spotlight.
One distinction that may be made is that some packages follow an integrated
approach that includes significant control of the program's main loop
\cite{Chombo,AMReX_JOSS}, while others relate to the mesh functionality as a
black-box add-on functionality
\cite{petsc-user-ref,BangerthBursteddeHeisterEtAl11}.
The latter thus would be most easily suited to link to standalone meshing
libraries, of which there are several of varying degrees of openness,
documentation, flexibility and parallel scalability.
For example, one package that has been around for a long time is Paramesh
\cite{paramesh}.

In this work we extend the \dune solver library \cite{dunereview:21} with
another grid interface to the open-source \pforest software
\cite{BursteddeWilcoxGhattas11}.
While \dune already supports about a dozen different mesh implementations
through its mesh interface \dunegrid, we undertake this new coupling effort
in order to inherit \pforest's practically unlimited MPI scalability
\cite{RudiMalossiIsaacEtAl15} as well as its relatively thin data
structures, and its native support for multi-block (forest) mesh topologies
in both 2D and 3D.

We choose a particularly narrow definition of the data exchanged between
\dunegrid and \pforest.
While both support full element connectivity, ghost exchange and 2:1 balance
across faces, edges, and corners, we only call upon the face components.
The remaining codimensions are generated by index sets inside \dune, which
supports face-based solvers like the finite volume method just as well as
fully connected finite element schemes.
This approach allows our interface to be adapted easily to further meshing
libraries that may only support face connections at the present time.

Most \dune solvers expect a 2:1 balanced mesh, that is, elements that
connect to neighbors that are same-, half-, or double-length, but no more
disparate.
We compare the native \pforest balance algorithm with a newly developed
iterative parallel scheme, which proves as a viable alternative when mesh
adaptation is frequent and incremental, i.\ e.\ non-recursive.
To remain focused on the face-connected perspective, we provide numerical
examples and performance results obtained from running \dune finite volume
solvers through the new \dunepfestgrid.



The remainder of this document is divided into a conceptual description of
our approach in \secref{concepts}, a review of existing and newly developed
parallel algorithms for this purpose in \secref{parallelalgorithms}, and
details on the implementation in \secref{implementation}.
We present numerical results on functionality and performance in
\secref{performancetesting}.

%% file: techdesign.tex
\section{Design concepts}
\seclab{concepts}

As a general principle in software engineering, combining packages that each
do one thing well keeps down technical debt and maintenance effort.
Any algorithmic and implementational intricacies and optimizations may be
confined within their respective package to remain invisible to the whole.
With a lean interface that fences complexity on the inside, the knowledge
required of an expert on one package to connect to any other remains
minimal.

What needs to be agreed upon however, and goes beyond and before specifying
a set of interface data types and functions, is a joint understanding of the
information that passes through them.
This information is ultimately defined by a set of mathematical concepts
that are independent of any software or realization.
Thus, we begin our development by exposing precisely the ideas, and not yet
the implementation, necessary for following the path of information through
the algorithmic flow of our design.

\subsection{The \dune grid interface}

The design and development of \dune started in late 2002,
with the main goal of \dune to
provide well-defined interfaces for the various components of a PDE solver
for which then specialized implementations can be provided.
\dune is not built upon one single grid data structure, nor is the intention
to focus on one specific discretization method only. All these components
are meant to be exchangeable.

As described in \cite{dunepaperII:08,dunereview:21}, the design principles of \dune are:
\begin{enumerate}
  \item  Keep a clear separation of data structures and
    algorithms by providing abstract interfaces that algorithms can be built
upon. Provide distinct, special purpose implementations of these data structures.
\item Employ generic programming using templates in
  C++ to remove any overhead of these abstract interfaces at
compile-time. This is very similar to the approach
    taken by the C++ standard template library (STL).
\item Do not reinvent the wheel. This approach allows us to reuse existing legacy code from our own and other
projects in one common platform.
    \label{enum:reusability}
\end{enumerate}

In particular item \ref{enum:reusability} is a very important
contribution to the actively discussed topic of \textit{research software sustainability}.


\subsubsection{Existing implementations of the \dune grid interface}

Currently, about ten different standalone implementations of the \dune grid
interface exist, providing various topologic and geometric representations
including general polyhedral meshes, different refinement techniques, and
other features.
Furthermore, about 13 so called \textit{meta grids} exist that enhance the
feature set of existing \dune grids.
An up-to-date list can be found on the project web
page\footnote{\url{https://dune-project.org/doc/grids/}}
or in
\cite{dunereview:21}.

Among those grid implementations only two, \dunealugrid~\cite{alugrid:16} and
\duneuggrid~\cite{Bastian1997} allow for parallel AMR. The core of both
implementations originates in the late 1990s, and data structures and
implementation choices made at the time are not really suitable for parallel
computers with large process counts.  For example, \dunealugrid can only
partition the coarsest mesh which means that computations on many processes
require very large macro meshes and thus do not allow for many levels of
additional adaptive refinement.

Here, \dunepfestgrid provides an alternative which allows distribution
of the workload at the element level based on a forests of octrees,
rendering it scalable to extremely high process counts.
Since \dunepfestgrid implements the \dune grid interface, it can be used as
a drop-in replacement for the quadrilateral and hexahedral meshes provided
by \dunealugrid or \duneuggrid without changing any other part of an
existing application code.

\subsection{A forest of octrees}
\seclab{forest}

Octree meshing certainly cannot be called a new technique anymore.
Trees have been a common concept in computer science from the very
beginning.  The understanding of recursivity as central to both trees and
space filling curves may go back more than a hundred years at least to
Lebesgue, and it resurfaces about half-way to the present in the paper by
Morton \cite{Morton66} on encoding geodesic information in two dimensions.

Leaping forward into the age of supercomputers, scalable octree codes
appeared that make use of a distributed octree data structure with
well-defined overlap between processes \cite{TuOHallaronGhattas05}.
Eliminating storage of tree edges and interior nodes, leaving only the
leaves in depth-first order leads to linear octrees employed at extreme
scales \cite{SundarSampathBiros08}.
These methods encode a parallel subdivision of the unit square or cube,
respectively, optionally mapped into simulation geometry.

The idea to tile a plane or volume using the images of multiple trees
has been developed alongside, but separately, where each tree would be
adaptively refined for a globally consistent non-uniform mesh.
The web of tree roots spanning the forest has been given a number of names,
for example the genesis mesh \cite{StewartEdwards04} or the coarse mesh
\cite{BangerthHartmannKanschat07, BursteddeHolke15}.
We will be using the term connectivity, as it is the convention used in the
\pforest quad-/octree library \cite{BursteddeWilcoxGhattas11} originally
conceived in 2007 and maintained to this day \cite{Burstedde25a}.

\subsubsection{A connectivity of trees}
\seclab{forestconnectivity}

There are various ways to define a connectivity of tree roots, where each
root is a topological hypercube.
(The following discussion applies as well to more general shape primitives
for roots and leaves, such as tetrahedra and prisms, which is the
fundamental idea of the \tetcode \cite{BursteddeHolke15,
HolkeBursteddeKnappEtAl23}.)
A prominent representation is a graph of nodes and edges, where each node is
a vertex of one or more connecting cubes.
Alternatively, the dual graph would refer to each tree as a node and connect
multiple nodes meeting at tree faces, and possibly tree edges and tree
corners.

The dual graph may be re-encoded by a set of lists that identify for each
tree the connections to others through all relevant codimensions, and the
group actions that define the relative orientation between neighboring
trees.
This design leads to an encoding devoid of graphs altogether, both for the
trees' connectivity and the linear list of leaves in depth-first order.
This is the way chosen by the \pforest and \tetcode software libraries.

In practice, the input to a simulation is frequently provided as a graph
derived from CAD descriptions or (coarse) mesh generators.
It may be necessary to convert this representation into the suitable set of
lists, which is in fact one functionality that we have added to the
new \dunepfestgrid grid interface described here.
Thus, we allow for the generic and portable family of mesh formats available
to \dune as well as calling builtin connectivity constructors for simple
shapes such as the cube, a brick of given proportions, a 2D or 3D cubed
sphere, and so on.

\subsubsection{Forest refinement and coarsening}
\seclab{forestadaptation}

Refinement is effected by the subdivision of any one element, and coarsening
is the reverse operation.
Both are controlled by an application.
Based on numerical results at any time in the flow of the simulation,
element-local indicators can be computed that flag an element to be refined
into the set of its children, or a complete family of children to be
coarsened into its parent element.

Indicators may be designed that flag any or all of the refine, keep, and
coarsen actions.
Thus, rules of precedence must be defined for a well defined process.
The rule we choose in accordance with common practice lets refinement take
precedence over keeping an element at the same size, and keeping takes
precedence over coarsening.
Furthermore, all children of a complete family of leaves must satisfy the
coarsening criterion to be eligible for coarsening as a group.

\subsubsection{Forest partitioning}
\seclab{forestpartition}

The space filling curve that arises naturally from any systematic recursive
ordering convention of a forest provides a global view of the mesh.
In this context, the order of the trees is a quality of the input data and
major to the linear ordering of leaves within any given tree.
A partitioning of the leaves between multiple processes defines a sub-view
for each process, where by our convention the sub-views are ordered
ascending by process number, disjoint and without gaps, and every sub-view
covers one (potentially empty) contiguous range of the global view.

Some forest implementations, such as \dune[ALUGrid] \cite{alugrid:16},
restrict the parallel partition to respect tree boundaries, limiting
scalability roughly to as many processes as there are trees.
Other implementations, such as \pforest \cite{BursteddeWilcoxGhattas11},
freely partition a tree between multiple processes to load balance the
global count over the whole forest, rendering its scaling properties
practically independent of the number of trees.


Refinement and coarsening operations are local, both to a process and to the
logic of the space filling curve: The children of an element refined are
ordered after the element's predecessor and before its successor.
Refinement and coarsening may change the number of leaves on any given
process by enlarging or shrinking its list of leaves, but preserve their
order and the shape of the partition boundaries in simulation space.

The process-local numbers of leaves $N_p$ can be equalized by repartitioning
as deemed necessary by allgathering it, computing $N = \sum_q N_q$ and
distributing the new leaf counts $N_p'$ equally between the $P$ processes.
Using the formula
\begin{equation}
  \eqnlab{partition}
  O_p' = \floors{ \frac{ p N }{P} }, \qquad
  N_p' = O_{p + 1}' - O_p'
\end{equation}
for any process $0 \le p < P$, it yields an optimal ($\pm 1$) target number
of leaves without further communication.
The leaves may be redistributed by a point-to-point messaging scheme with a
well-defined handshake (sender and receiver ranks and message sizes are
computable from $(N_q)$, $(N_q')$ alone).
This scheme is so fast \cite{BursteddeHolke16b} that it would be a mistake
\emph{not} to repartition the mesh if it will improve the load balance, even
slightly.
The formula and algorithm has been generalized by weighting each leaf for
equi-partitioning the weights.

With the advent of linear octree methods, it becomes realistic to achieve
partition-independent meshing.
This means that the global mesh evolves invariant under the number of
processes $P$ and the precise partitioning formulas and weights used.
One problem that had to be solved along the way is to not prevent coarsening
by splitting a family of elements across a process boundary.
This modification has been implemented by a post-processing step to
\eqnref{partition} that involves at most two short messages per process.

One more building block to partition independence is to load and store the
mesh to disk in a serial-equivalent format, which is one functionality of
the \pforest software.

\subsubsection{Interrogating the forest mesh}
\seclab{forestinterrogation}

The simplest and in fact a reasonably effective way to iterate through the
mesh is to loop over the process-local trees, that is, those that contain at
least one local element, and in each tree over its local elements.
Since the tree number takes precedence over the element number in a tree,
the local view of the forest is implicitly traversed depth-first.
This linear iteration is sufficient for computing adaptation indicators,
weights for partitioning, and to read and write data to other formats.

In addition to a volume loop, any element-based numerical method requires
reading data from nearby elements.
These elements may be direct neighbors through faces only, as is the case
for finite volumes and discontinuous Galerkin methods, or through edges and
corners as well, as is the case for more elaborate finite volume and finite
element schemes.
Some methods even require information from next-nearest neighbors.
Thus, we need a way to enumerate all relevant local element neighbors.

Exploiting the total order of the leaves, for any given query element we may
construct hypothetical element neighbors and binary search for them in the
ordered list of leaves \cite{BursteddeWilcoxGhattas11}.
An optimization to diminish the logarithmic factor in the above approach
lies in orchestrating a top-down traversal of the local forest that
maintains the neighbor context as an invariant.
Its speed improves due to narrowing the length of search windows as we
descend to finer levels, and by caching and reusing search results.
If a ghost layer of direct remote-process neighbors is known, which is
another sorted (albeit non-contiguous) list of leaves, it can be seemlessly
integrated into the traversal for a complete parallel neighborhood context.

Since the recursion can be equivalently reformulated as a while loop, and it
maintains strict invariants related to traversing the tree in ascending
order of the leaves, we call this process the mesh iterator
\cite{IsaacBursteddeWilcoxEtAl15}.
Naturally, separate mesh iterators exist on various levels of the software
stack, so in addition to this neighbor-aware forest iterator we will
encounter a high-level neighbor iterator in the \dune library.
We will detail below how the latter can be implemented by calling the
former.

\subsection{On 2:1 mesh balance}
\seclab{forestbalance}

We refer to the 2:1 balance of a mesh as the property that direct neighbor
elements must not differ in length by more than a factor of two.
Such a property serves as a simplifying constraint on the otherwise
arbitrary grading of the mesh.
2:1 balancing limits the number of cases that can occur between neighbor
elements and thus the complexity of any numerical method.
Depending on the topological demands of the method, 2:1 balancing may be
asked for across faces, or also across edges, and across corners as the
strictest form.

Most numerical solvers, including those inside \dune, will theoretically
admit a formulation on non-balanced meshes, but the complexity of
implementing this variability would often exceed the complexity of adding a
2:1 balancing algorithm to the mesh logic.
When a local mesh neighbor iteration is available, for example in the form
discussed in \secref{forestinterrogation}, the so-called \emph{ripple}
balance can be implemented: Each element exchanges its refinement level and
flags with its direct neighbors, in parallel invoking one round of messages,
to decide on its own target refinement level.
Then the mesh is refined synchronously and we repeat the process for
multiple rounds until the mesh eventually stabilizes.
This approach has many names, for example diffusive mesh smoothing, and was
employed and demonstrated successfully at scale by the Octor code
\cite{TuOHallaronGhattas05} and taken up again in the \tetcode
\cite{HolkeKnappBurstedde21}.

The state of the art of 2:1 balancing has since shifted towards
non-iterative algorithms.
Based on the concept of the insulation layer \cite{SundarSampathBiros08}, it
is possible to formulate one round of point-to-point communication and
combine it with an ensuing merge-sort and unification, which yields
scalability to over 1e5 CPU cores \cite{BursteddeWilcoxGhattas11}.
An optimization to reduce the exchange to a minimal set of seed quadrants
further lowers the memory footprint and run time of this method
\cite{IsaacBursteddeGhattas12}.
Finally, synchronization can be eliminated further by finely tuned encoding
and sorting algorithms \cite{SuhIsaac20}.

To keep this paper focused, we choose only solvers and numerical examples
that use 2:1 balance.
The modularity of our software interface allows us to exercise both types of
algorithms, the one-shot black-box algorithm provided by \pforest and an
iterative ripple procedure leveraging the \dune mesh data structure.
While the one-shot algorithm is preferred for meshes of unknown grading, the
ripple algorithm requires very few rounds when invoked after every
refinement and coarsening in an incremental workflow.
We will provide algorithmic details in the following sections.

\section{Parallel support algorithms}
\seclab{parallelalgorithms}

The \dunegrid interface requires specific well-defined mesh information to
be provided by the grid backend, in our case \pforest.
One part is covered by iterating over each mesh face on the local process,
together with the element indices and orientations on both of its sides.
Another part is supported by the face ghost exchange, where designated data
on each local element at a process boundary is communicated to every
face-adjacent process.
For the \dunepfestgrid interface, we limit all adjacency information to face
connections, keeping the data passed through the interface at a minimum.
We accept the tradeoff that some information necessary for mesh balancing
must hop diagonally between elements even for face-only balance, which
requires $d$ subsequent face hops.
This we must be prepared to iterate any ripple balance algorithm until
completion.

\subsection{A mesh face iterator}
\seclab{pforestfaceiterator}

The \pforest library provides a mesh iterator that enumerates all
process-local faces, edges, and corners, together with relevant data on the
elements surrounding a particular mesh boundary object
\cite{IsaacBursteddeWilcoxEtAl15}.
Presently, this algorithm is restricted to 2:1 balanced meshes, which is
fine for the scope of this submission.
Its restriction to face connections suffices to populate the intersections
of the \dunegrid.
We have developed an experimental non-balanced version, which prepares us to
extend our method to non-2:1-balanced meshes whenever so desired.

\subsection{The ghost layer}
\seclab{pforestghostlayer}

In \pforest, we have the option to populate a ghost layer of all direct
off-process element neighbors using one symmetric round of communication.
It can be configured to gather face neighbors only, or additionally edge
neighbors in 3D, or additionally corner neighbors.
For the present face-oriented interface to \dunegrid, the face-only version
suffices.
It can be constructed once after every new mesh adaptation and partitioning
and subsequently be accessed read-only any number of times.

The ghost layer of \pforest comes with an exchange algorithm, which allows
for transfering designated data on the local elements of one process to all
others that require a share.
Since the ghost elements follow the space filling curve like the real ones,
all ghosts from one process are ordered after all ghosts from
lower-numbered, and before all ghosts from higher-numbered processes.
Thus, the exchange algorithm requires no handshake and operates with one
aggregated and overlapped point-to-point exchange per pair of adjacent
processes \cite{MirzadehGuittetBursteddeEtAl16}.

The \pforest ghost exchange algorithm is thus highly efficient and allows
for implementing e.g.\ parallel mesh-smoothing and balance marking
algorithms without invoking the comparably heavier general-purpose mesh
iterator introduced just above.

\subsection{An iterative 2:1 balance algorithm}
\seclab{iterativebalance}

In this section we present an iterative ripple algorithm to adjust a marking
provided by the user such that the resulting adapted grid fulfills the
constraint of a 2:1 balance across all element faces.
This algorithm serves as an alternative to the monolithic \pforest mesh
balance, whose latest incarnation is described in \cite{SuhIsaac20}.

We follow the notation from \cite[Ch.\ 2]{kloefkorn:phd} for a hierarchic
grid $\hgrid$ with leaf elements $\grid^0$ that fulfills the 2:1 balance
constraint (as each grid in \dune satisfies this property on creation).
We denote a marking $\mathcal{M} := \{ m_{\elem} : m_{\elem} \in \{-1, 0,
1\}, \elem \in \grid^0 \}$ specified by the user such that
$m_{\elem} = -1$ indicates coarsening of $\elem$, $m_{\elem} = 0$ keeping
$\elem$ as is, and $m_{\elem} = 1$ refinement of $\elem$.

The general logic is that an element $\elem \in \grid$ that is marked for
refinement ($m_{\elem} = 1$) will be refined in any case, whereas elements
that are marked for coarsening ($m_{\elem} = -1$) or keeping ($m_{\elem}=0$)
might still be refined to satisfy the 2:1 balance constraint.
For example, neighboring elements will be marked for refinement if an
element that is marked for refinement already presents a level difference
such that the neighbors are on a coarser level.
The detailed procedure is explained in Algorithm \ref{alg:balancedmarking}.

In \dunepfestgrid changing the marking of elements is cheap to compute. The only
drawback of the algorithm is the necessity of synchronization of the element
marking between processes and the global synchronization whether consistent
marking has been reached on all.  In a worst case scenario, $P$ steps of the
algorithm would have to be carried out when running on $P$ processes, and
even a multiple taking into account diagonal hops.
To prevent such situations, Algorithm \ref{alg:balancedmarking} terminates
after at most $3$ unsuccessful steps and the monolithic \code{p4est_balance}
is called as a fallback to produce a 2:1 balanced grid with certainty
\cite{IsaacBursteddeGhattas12, SuhIsaac20}.
For the experiments carried out in this article the Algorithm
\ref{alg:balancedmarking} always terminated within the prescribed limit and
\code{p4est_balance} was never called.

In the current implementation an possible overlap of synchronization of the element
marking between processes (Line 2 in Algorithm \ref{alg:balancedmarking})
and modification of the marking locally has not been
taken into account but would certainly improve the scalability of this
algorithm.

\begin{algorithm}
  \scriptsize
  \DontPrintSemicolon
  \caption{Balanced marking}
  \label{alg:balancedmarking}
  \tcc{Repeat up to 3 times}
  \For{$i = 0,1,2$}
  {
    \tcc{Communicate marking from local process boundary to ghost cells}
    \pforestexchange( $\mathcal{M}$ )

    \tcc{Until locally consistent, repeat }
    $r_p = 1$ \\
    \While{ $c = 0$ }
    {
      $c = 1$ \\
    \tcc{For all leaf elements $\elem \in \grid$}
      \For{$\elem \in \grid^0_p$}
      {
        \tcc{If element $\elem$ is marked for refinement}
        \If{$m_{\elem} = 1$}
        {
          \For{$\neig \in \interset$}
          {
            \If{$l_{\neig} < l_{\elem}$ \LAND $m_{\neig} < 1$}
            {
              \tcc{Mark neighbor for refinement}
              $m_{\neig} = 1$; $c = 0$
            }

            \If{$l_{\neig} < l_{\elem}$ \LAND $m_{\neig} < 0$}
            {
              \tcc{Leave neighbor as is}
              $m_{\neig} = 0$; $c = 0$
            }
          }
        } 

        \tcc{If element $\elem$ is marked to left as is}
        \If{$m_{\elem} = 0$}
        {
          \For{$\neig \in \interset$}
          {
            \If{$l_{\neig} > l_{\elem}$ \LAND $m_{\neig} = 1$}
            {
              \tcc{Also mark element for refinement}
              $m_{\elem} = 1$; $c = 0$
            }
          }
        } 

        \tcc{If element $\elem$ is marked for coarsening}
        \If{$m_{\elem} = -1$}
        {
          \For{$\neig \in \interset$}
          {
            \If{$l_{\neig} > l_{\elem}$}
            {
              \If{$m_{\neig} > 0$}
              {
                \tcc{Mark for refinement if neighbor is marked}
                $m_{\elem} = 1$; $c = 0$
              }
              \Else{
                \tcc{Keep element since neighbor is finer}
                $m_{\elem} = 0$; $c = 0$
              }
            }

            \If{$l_{\neig} = l_{\elem}$ \LAND $m_{\neig} > 0$}
            {
              \tcc{Keep element since coarsening would unbalance }
              $m_{\elem} = 0$; $c = 0$
            }
          }
        }
      }
      \tcc{If any marking was altered then the state is unresolved}
      \If{$c = 0$}
      {
        $r_p=0$
      }
    } 

    \tcc{Allreduce the resolved flags over all processes}
    \If{ $\min_{p \in P} r_p = 1$ }
    {
      \Return
    }
  } 

  \tcc{If no resolved state was reached, fall back to monolithic balance}
  \pforestbalance( $\hgrid$ )
\end{algorithm}


%% file: implementation.tex
\section{Implementation}
\seclab{implementation}

After the preceding discussion from a predominantly conceptual perspective,
we move on to detailing important aspects of the concrete implementation of
the \dunepfestgrid.
Part of it amounts to augmenting the \pforest software with additional entry
points, and part amounts to providing a new instantiation of the \dunegrid
interface that internally calls upon a small set of \pforest functions.

\subsection{Interfacing to \pforest}

%
%

The present development is based on the latest public releases of \pforest
\cite{BursteddeGriesbachBrandtEtAl25} and its support library \libsc
\cite{BursteddeGriesbachHirschEtAl25}.
As hinted to earlier, we extend the mesh iterator function with entry points
that allow for unbalanced meshes.
Another addition is to generate a \pforest connectivity object (that
connects individual trees topologically into a forest) from any coarse
\dunegrid provided as input.
Finally, we have added an algorithm to unify the representation of mesh
coordinates on the boundary of touching trees.
The parallel algorithm to collect the ghost layer, as well as the ghost
exchange functionality, and the high-level calls to initialize, refine,
coarsen, and repartition a forest mesh, are used without modification.


For this paper, we are including an alternative design of the \pforest
highlevel algorithms and parallel setup that is targeted to shared memory
computation.
In particular, we are replacing MPI communication with read and write
accesses to MPI-3 shared memory where possible.
This branch of the code is tentatively named \pforthree and considered
experimental until further notice.
One key difference to the standard \pforest design is that it presently
operates without 2:1 balance.

While the standard \pforest encourages to populate a ghost layer before
passing it as context to the mesh iterator, in \pforthree we are able to
construct the iterator without referring to a ghost structure, since
off-process neighbor quadrants are available in MPI-3 shared memory.
To support the call to \pforestexchange by the \dunepfestgrid, we populate a
standard \pforestghost object by running through the \pforthree iterator and
collecting the required data.
Details are listed in Algorithm~\ref{alg:pforthreeghost}.
\begin{algorithm}
  \scriptsize
  \DontPrintSemicolon
  \SetAlgoLined
  \SetKwInOut{Input}{Input}
  \SetKwInOut{Output}{Output}

  \caption{Populate \pforestghost from shared-memory \pforthree data}
  \label{alg:pforthreeghost}%

  \Input{\pforthree object $\mathcal{P}$}
  \Output{\pforestghost object $\mathcal{G}$}
  \BlankLine


  \BlankLine
  \tcc{For all local faces $f \in \mathcal{F}$ of the forest $\mathcal{P}$}
  \For{$f \in \mathcal{F}$}
  {
    \tcc{If $f$ connects sides $s_g$ and $s_m$ from different processes}
    \If{$s_g \not\in p_l$ \LAND $s_m \in p_l$}
    {
      $p_r \gets$ Binary search to find remote owner of $s_g$\;
      \tcc{Create hash key $k_g$ for hash table $\mathcal{H}_g$}
      \If{$k_g = (s_g,\ p_r) \not\in \mathcal{H}_g$}
      {
        $\mathcal{H}_g \gets \mathcal{H}_g \cup k_g$\;
        $\mathcal{G}_g \gets \mathcal{G}_g \cup s_g$\;
      }
      \tcc{Create hash keys $k_m$ and $k_m^{p_r}$ for hash table $\mathcal{H}_m$}
      \If{$k_m = (s_m,\ p_l) \not\in \mathcal{H}_m$}
      {
        \tcc{Add $k_m$ with key-order index $i_m$ in $\mathcal{H}_m$}
        $\mathcal{H}_m \gets \mathcal{H}_m \cup \{k_m,\ i_m\}$\;
        $\mathcal{G}_m \gets \mathcal{G}_m \cup s_m$\;
      }

      \If{$k_m^{p_r} = (s_m,\ p_r) \not\in \mathcal{H}_m$}
      {
        $\mathcal{H}_m \gets \mathcal{H}_m \cup \{k_m^{p_r},\ i_m\}$\;
        \tcc{Push back $i_m$ in per-processor array of permutations}
        $M[p_r] \gets M[p_r] \cup i_m$
      }
    }
  }
  \tcc{Sort ghost $\mathcal{G}_g$ and mirrors $\mathcal{G}_m$ arrays in piggy3
  order}
  Sort $\mathcal{G}_g$\;
  Sort $\mathcal{G}_m$, update $M$ arrays accordingly\;

  \tcc{Merge per-process mirror arrays}
  Merge $M[0..p-1]$ into auxiliary proc$\leftrightarrow$mirrors indices array\;

  \For{$g \in \mathcal{G}$}
  {
    \tcc{Build mapping $\mathcal{M}: \text{global index}\ I_g \mapsto
    \text{position}\ i_{\mathcal{G}}$ in $\mathcal{G}_q$}
    $\mathcal{M} \gets \mathcal{M} \cup (I_g, i_{\mathcal{G}})$
  }

\end{algorithm}

\subsection{Grid and entities}

\dunepfestgrid implements a \dune grid by storing an object for connecting
the trees as \pforestconnectivity, the forest containing the local leaves as
\pforestforest and a corresponding ghost layer \pforestghost.

The connectivity object stores the macro grid (connected trees of \pforest)
and stays fixed over the course of the simulation.
The forest of octrees provides an ordered collection of \pforestquadrant{}s
that form the leaf elements of the grid in the \dune sense and permits
dynamic adaptation and repartitioning.

The information provided by a leaf \pforestquadrant is augmented in \dune by
the \code{EntityInfo} object which stores the quadrant and some supporting
information. The grid then stores a \code{std::vector} containing the
\code{EntityInfo} for each leaf element of the grid. The memory consumption of
this object is $24$ bytes. Elements that are owned by a process
(\code{interior}) are stored in
the front part of this vector and \code{ghost} elements are stored in the rear
part of this vector.


\subsubsection{Iterators}

\dunepfestgrid provides iterators for the leaf grid through the
\code{LeafGridView}.
For the elements $\elem^0 \in \grid^0$ the iterator simply iterates through
the element vector.
Since the \code{interior} and \code{ghost} elements are stored in
consecutive chunks these iterators can be implemented very efficiently.
Iterators for higher codimension entities are constructed using the element
iterator and a boolean vector to only visit each entity once.

\subsubsection{P4estEntityInfo}

\code{P4estEntityInfo} implements the essential information needed to describe an
entity in the grid. This class stores
\begin{itemize}
  \item \pforestquadrant a pointer to a \pforestquadrant data structure ($8$
bytes),
  \item \code{treeid}, the index of the tree or macro element in \dune notation ($4$ bytes),
  \item \code{quadid}, the local Morton curve index of the quadrant ($4$ bytes),
  \item \code{persistentIndex}, the local persistent index which is unique per codim ($4$ bytes),
  \item \code{codim}, the codimension of the entity ($1$ byte),
  \item \code{subEntity}, the sub entity number inside the codimension $0$ entity ($1$ bytes),
  \item \code{flags}, a collections of bit flags to store information such as
    \code{isLeaf}, \code{isGhost}, or \code{isNew} ($1$ byte in total).
\end{itemize}
In total the \code{P4estEntityInfo} object requires $24$ bytes in memory. With this
object each entity in the grid can be uniquely identified locally on a process.

\subsubsection{P4estGridEntity and P4estGridEntitySeed}

The \code{P4estGridEntity} implements the class providing elements ($c = 0$), edges ($c = d-1$), faces ($c=1$)
and vertices ($c=d$) of the grid class. The entity class stores an object of
type \code{P4estEntityInfo} which provides the topological information. In
addition a pointer to the grid class is stored and a geometry object providing
the geometrical representation of the entity.
In total the entity object requires $40$ bytes memory consumption.
The \code{P4estGridEntitySeed} class simply stores a \code{P4estEntityInfo}. The
only difference is that the \code{codimension} is a template parameter to allow
for construction of the entity class which also contains the \code{codimension} as template parameter.

\subsubsection{P4estGridGeometry}

The implementation of the geometry class follows more or less exactly the
implementation of the geometry class of \dunealugrid for cuboid elements.

The geometric coordinates for each element are provided by \pforest through the
function \code{p4est_qcoord_to_vertex}. This function is $\mathcal{O}(1)$
but relatively costly and therefore the geometrical vertex coordinates of
each element are computed only when an element is created and then stored in
a \code{std::vector} structure which is accessed using the persistent index
of an element.

\subsection{IdSet and IndexSets}

\dunepfestgrid provides index maps (\code{P4estGridIdSet}) which contain the
globally unique \code{P4estEntityId} and persistent
(\code{P4estGridPersistentIndexSet}) and consecutive
(\code{DefaultIndexSet}) index sets.

\subsubsection{GlobalId and GlobalIdSet}

To uniquely identify entities across processes and
codimensions \dunepfestgrid implements the \code{P4estEntityId}.

This structure stores $d+1$ \code{qcoord_t} which are currently \code{int32_t}
in \pforest. For entities of higher codimension this is used to
store the coordinates of the entity's center inside a tree and the $d+1$st
integer is used to encode the \code{treeId} and \code{codimenson}.
For entities of codimension $0$ (elements) the globally unique element number is
used as global id.

To compute the coordinates of the center of an entity the following new convenience
functions are provided by \pforest:

\begin{itemize}
  \item \code{p4est_quadrant_corner_coordinates} for vertices,
  \item \code{p4est_quadrant_face_coordinates} for edges in 2D and faces in 3D,
  \item \code{p8est_quadrant_edge_coordinates} for edges in 3D, and
  \item \code{p4est_quadrant_volume_coordinates} for volumes in 2D and 3D.
\end{itemize}

\begin{figure}
  \subfloat[$4$ trees]{\label{fig:trees}
  \includegraphics[width=0.45\textwidth]{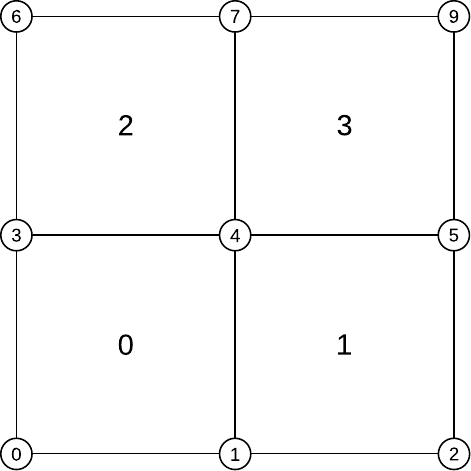}} \hfill
  \subfloat[refined]{\label{fig:trees_refined}
  \includegraphics[width=0.45\textwidth]{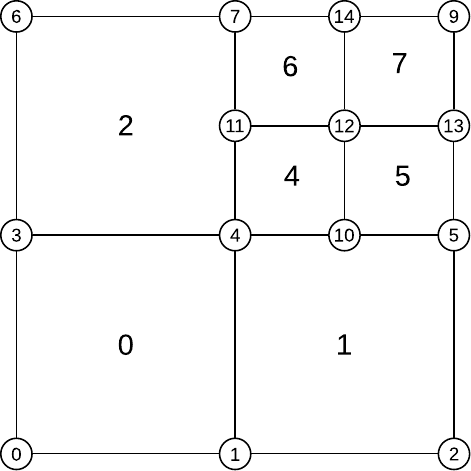}}
  \caption{Macro grid (left) and grid where element $3$ has been refined once
  (right).}
  \label{fig:ids}
\end{figure}

These functions compute the unique coordinates of the mid point of an object
inside a tree of the forest. These coordinates are unique inside a tree but
need to be canonicalized across tree boundaries which is achieved by the
function \code{p4est_connectivity_coordinates_canonicalize}.
For a grid $\grid$ with $\mathrm{dim}(\grid) = d$ the
id (\code{P4estEntityId}) of an entity $\elem^c \in \grid^c$
of an entity is computed as a tuple of $d+1$ 32-bit integers
\begin{equation}
  \id(\elem^c) := \big (\mathbf{x}_0, ...,\mathbf{x}_{d-1}, t_{id} (d+1) + c \big),
\end{equation}
where $0 \leq \mathbf{x}_i \leq 2^{30}$ for $i =0,...,d-1$
are the unique coordinates (\code{p4est_qcoord_t})
of the mid point of $\elem^c$ inside the uniformly refined tree with level $30$,
$t_{id}$ is the tree id, $d$ the dimension of the grid and $c$ the
codimension of the entity. For entities at the tree boundary either one or
several or even all coordinates are $0$ or $2^{30}$.
In Figure
\ref{fig:trees} a unrefined forest with $4$ trees is displayed. In this grid,
for example, vertex $0$ has the global id $(0,0,2)$.
Vertex $9$ has the global id $(2^{30},2^{30},11)$. For vertices at tree
boundaries, such as vertices $1,3,4,5$ and $7$ the tree local \code{p4est_qcoord_t} is
canonicalized using the \code{p4est_connectivity_coordinates_canonicalize} to
obtain the unique coordinates inside the tree with the smallest tree id. For
example, vertex $4$ has the global id $(2^{30}, 2^{30}, 2)$ which is the
tree local coordinate of vertex $4$ inside tree $0$.
For refined grids, for example as in Figure \ref{fig:trees_refined}, vertices
and other objects that are located inside a tree such as vertex $12$ have the
global id given by the \code{p4est_qcoord_t} and the tree id and codimension.
For vertex $12$ we have the global id $(2^{29}, 2^{29}, 11)$.

The overall memory footprint of an \code{P4estEntityId} is $128$ bit for
both, 2D and 3D versions.
The global id of each entity is accessed through the
\code{GlobalIdSet} provided by the grid. The \code{LocalIdSet} is just an alias
and the same implementation is used.

In the implementation of \dunepfestgrid the global ids are used to generate
persistent indices for higher codimension entities and ensure that entities
shared by multiple elements have the same persistent index. These indices are
then stored in an element-by-element fashion to ensure efficient access.

\subsubsection{Persistent index}

Using the global ids a persistent index is implemented and stored in an
element-by-element fashion. The persistent index implements a \dune index set
with a deviation from the interface by allowing a non-consecutive index set.
Each entity keeps the same process local persistent indices as long the entity
exists. Change can only occur through grid adaptation or load balancing.

The concept of persistent indices is detailed in \cite[Def 2.5.9]{kloefkorn:phd}.
The persistent index set is used to implement the \code{PersistentContainer}
which attaches data to grid entities that is preserved throughout grid
modification. The same concept is implemented in \dunealugrid.

\subsubsection{IndexSets}

Following \cite[Def 2.2.13]{kloefkorn:phd} all other index sets are consecutive index sets
and the implementations are based on the persistent index and reuse the
\code{DefaultIndexSet} provided by \dunealugrid which uses the
\code{PersistentContainer} to create consecutive indices for a given grid view.
Details are given in \cite{dunereview:21, kloefkorn:phd}. The current  \dunepfestgrid
only provides a \code{LeafGridView} and therefore only an index set for the leaf
grid.

\subsection{Intersections}

A \dune intersection between two elements (codim $0$ entities) is computed by
the \code{p4est_dune_iterate_balanced} function provided by the \pforest library.

This iterator visits each intersection only once providing left and right neighbors
and local face numbers which is then used to setup a \code{std::vector} structure.
For each element we then store a fixed length vector with at most $d 2^{d}$ entries
containing all neighboring
element indices as well as corresponding local face numbers.
For intersections with the boundary a negative neighbor index and the local face
number of the inside element is stored.
This then allows to efficiently create an intersection iterator
for each element only involving iteration of a fixed length array of integers.

The setup of this structure also allows the efficient implementation of the
balanced marking strategy described in Algorithm \ref{alg:balancedmarking}.

\subsection{Python bindings}

By implementing the \dune grid interface \dunepfestgrid automatically also comes
with Python bindings for the grid.
The code

\begin{figure}
  \begin{center}
    \subfloat[Cartesian grid of 256 trees]{\includegraphics[width=0.45\textwidth]{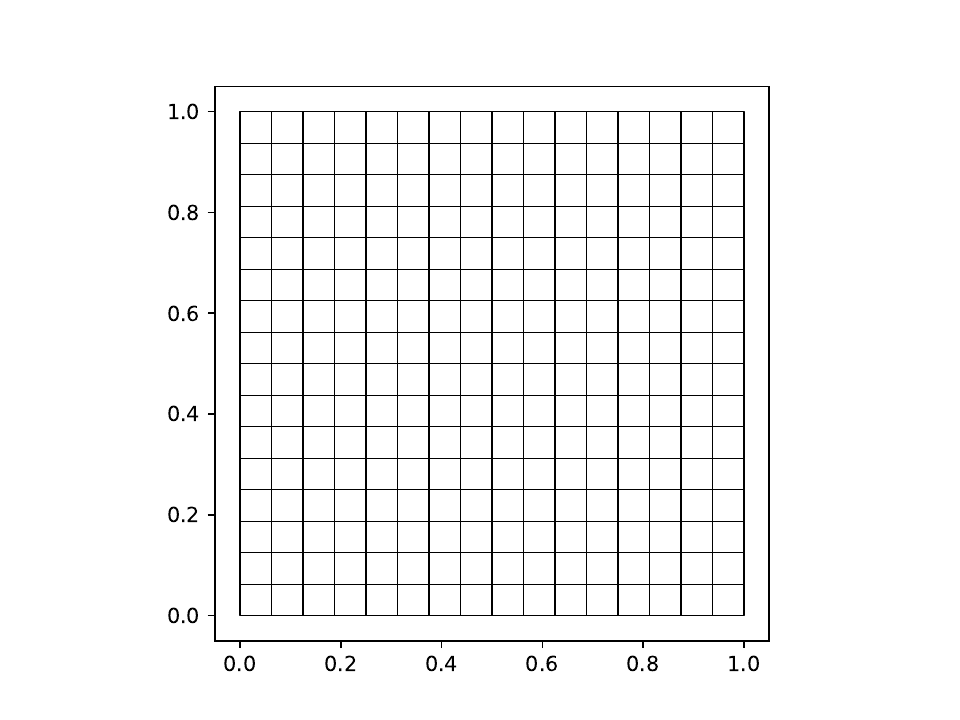}}
    \subfloat[Unstructured grid of 3 trees]{\includegraphics[width=0.45\textwidth]{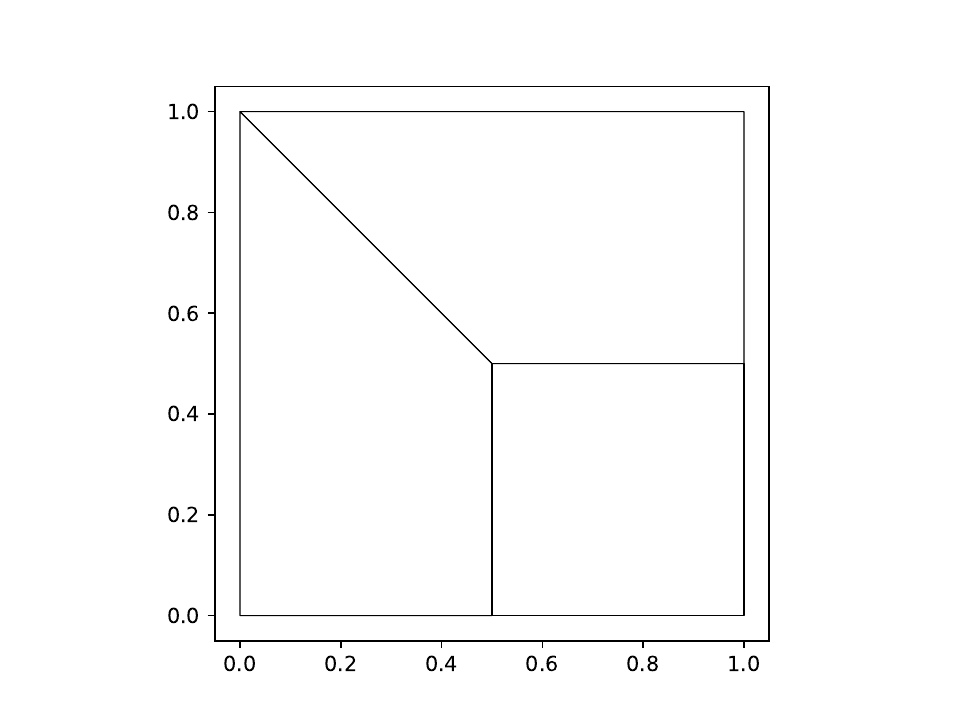}}
    \caption{A Cartesian (a) and an unstructured coarse grid (b) created
             using the Python interface to \dunepfestgrid.}
    \label{fig:pythongrids}
  \end{center}
\end{figure}

\newcommand{\picturepath}{pictures/grid}
\begin{python}[p4estdemo.py]
from dune.p4estgrid import p4estCubeGrid
from dune.grid import cartesianDomain

# create Cartesian grid
domain = cartesianDomain([0, 0], [1, 1], [16, 16])
gridView = p4estCubeGrid(domain)
gridView.plot()

# create unstructured grid
mesh = { "vertices":[ [ 0, 0], [0.5, 0],
                      [ 1, 0], [0.5, 0.5],
                      [ 1, 1], [  1, 0.5],
                      [ 0, 1 ] ],
      "cubes": [ [0, 1, 6, 3], [1, 2, 3, 5], [ 3, 5, 6, 4] ] }
gridView = p4estCubeGrid( mesh )
gridView.plot()
\end{python}
creates a \dunepfestgrid for the unit quadrilateral with $16^2$ trees and an
unstructured grid (see Figure \ref{fig:pythongrids}).
\dunepfestgrid can, for example, be used as a seamless replacement for
\dunealugrid.

\subsection{Derivations from the \dune grid interface}

In this implementation of the \dune grid interface the deliberate decision was
made to not support the full grid hierarchy but only the finest level of the
grid: the leaf level. Therefore, only the \code{LeafGridView} is available and
hierarchic information such as \code{entity.father()} cannot be accessed. For
adaptation only the callback version \code{grid.adapt( dataHandle )} is
available. During the adaptation process the methods of \code{dataHandle} will
be called with appropriate parent-child element pairs. All methods related to
level-wise interaction with the grid do not work, e.g.\ \code{LevelGridView} is
not available.


%% file: numericalexperiments.tex

\input{perfmeassures.tex}

\subsection{Rotating doughnut experiment}
\label{sec:rotatingball}

\renewcommand{\picturepath}{results/ball}

\begin{figure}
  \begin{center}
  \subfloat[$t=0$]{\includegraphics[width=0.3\textwidth]{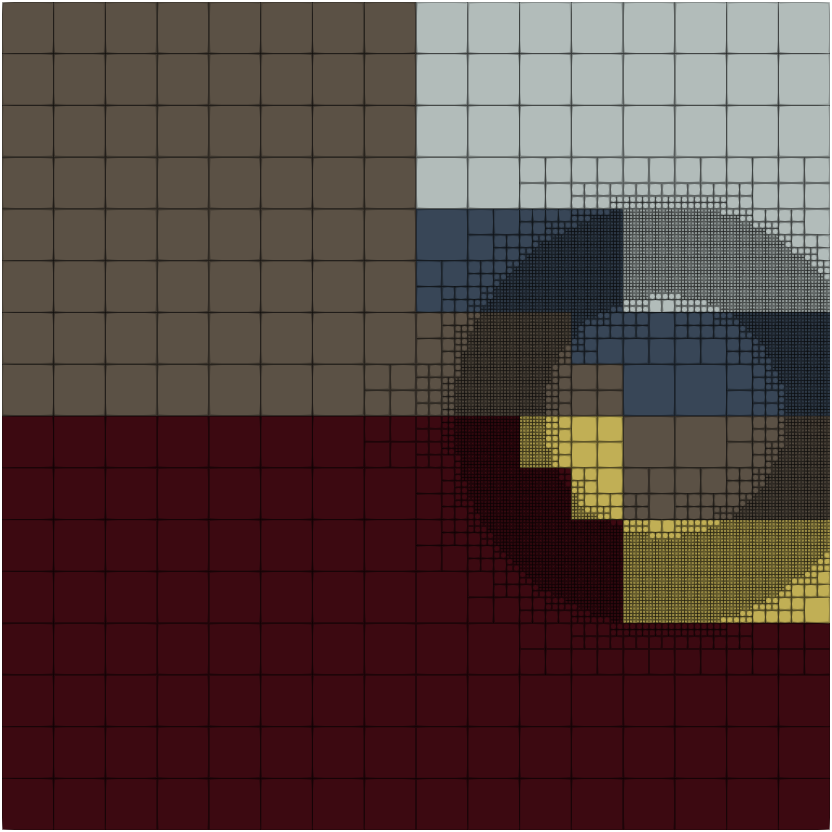}} \hfill
  \subfloat[$t=0.7$]{\includegraphics[width=0.3\textwidth]{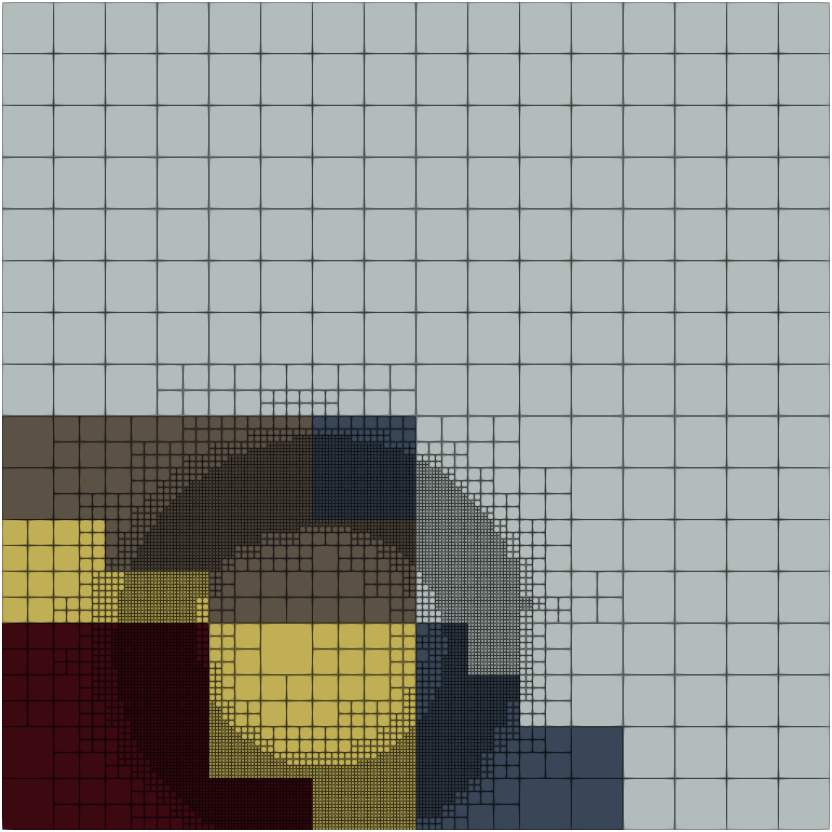}} \hfill
  \subfloat[$t=0.15$]{\includegraphics[width=0.3\textwidth]{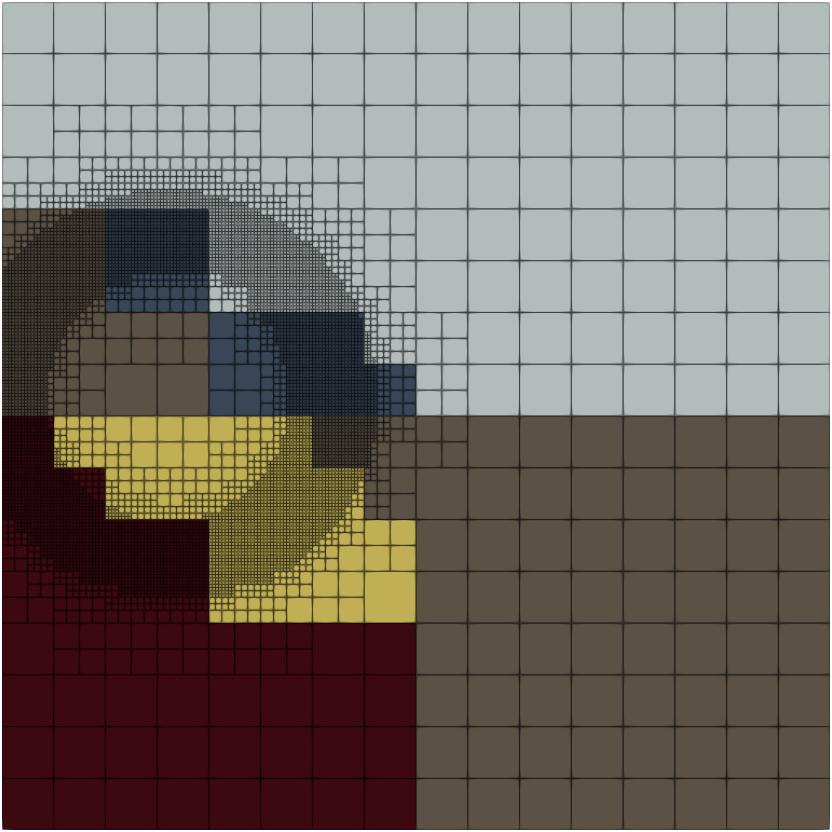}} \\
  \subfloat[$t=0$]{\includegraphics[width=0.3\textwidth]{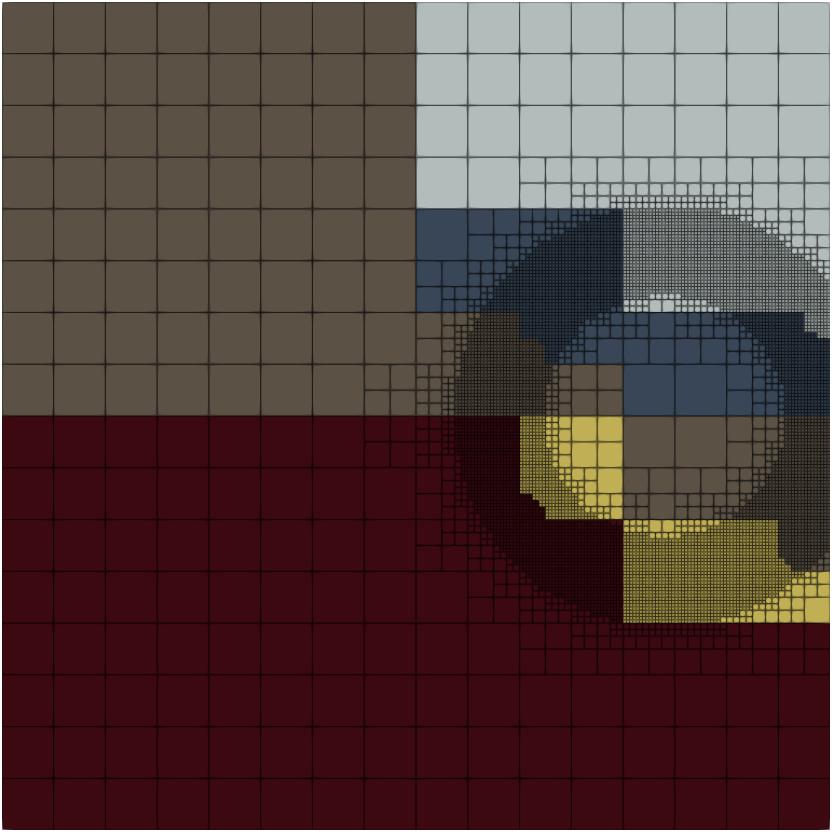}} \hfill
  \subfloat[$t=0.7$]{\includegraphics[width=0.3\textwidth]{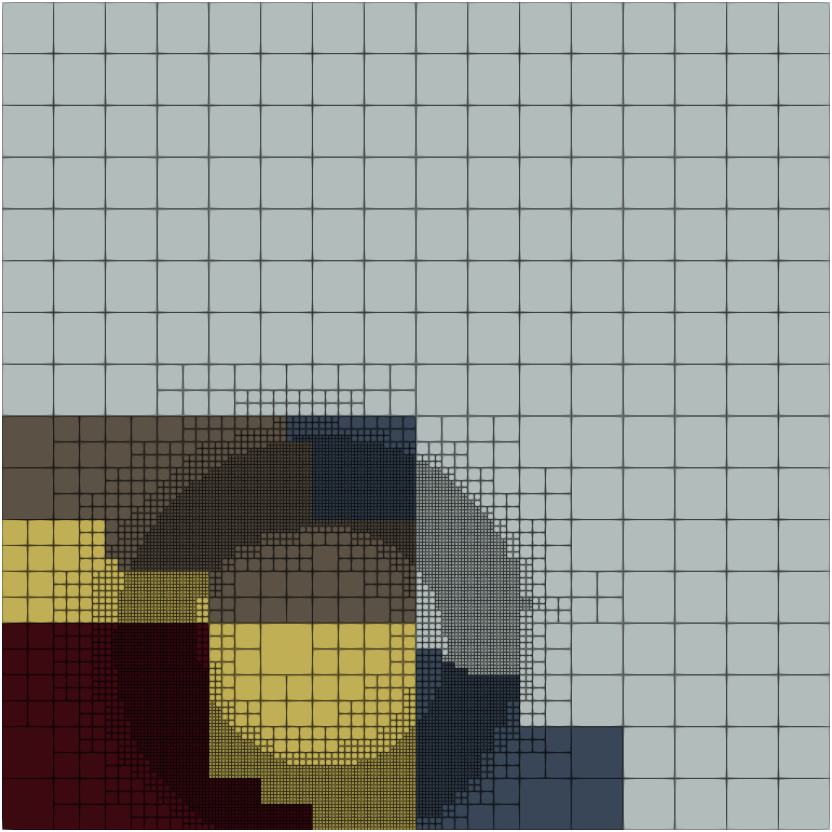}} \hfill
  \subfloat[$t=0.15$]{\includegraphics[width=0.3\textwidth]{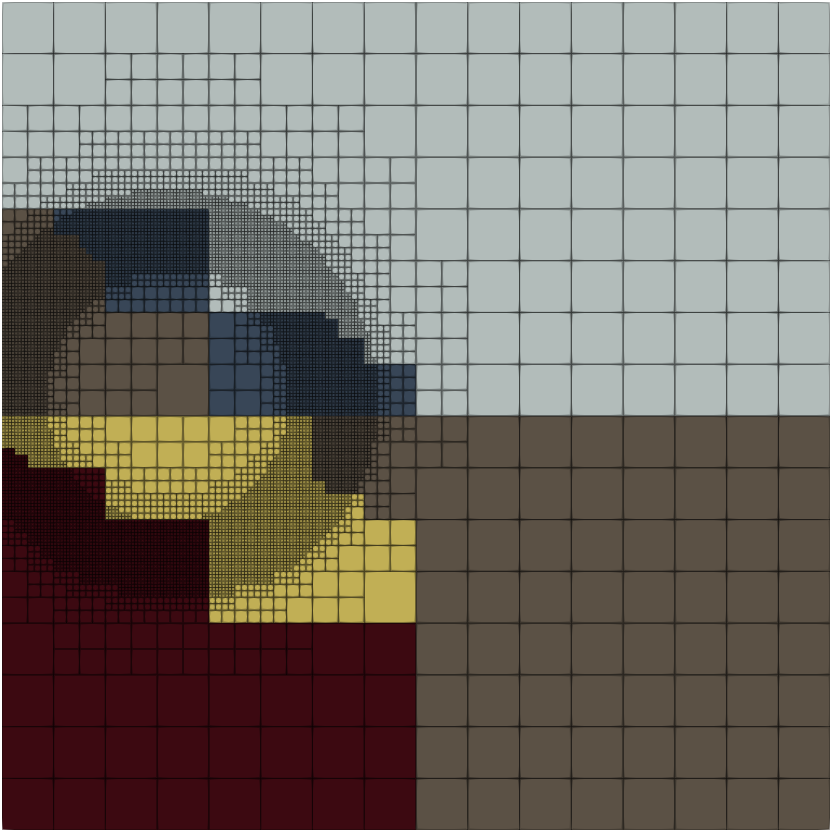}}
  \caption{Rotating doughnut experiment using
    \dunealugrid (top row (a)--(c)) and \dunepfestgrid (bottom row (d)--(f))
    at 3 different time points.
  The initial macro
  grid has $16\times16$ cells and a total refinement depth of $4$ levels was
  allowed for this run. The computation was carried out on $5$ cores and the
  different partitions are indicated by the different colors. A difference
  between the partitioning with \dunealugrid compared to \dunepfestgrid can be
  clearly seen. The partitioning with \dunealugrid is between the macro
  cells, while \dunepfestgrid partitions to the individual element.}
    \label{fig:ball_pics}
  \end{center}
\end{figure}

\begin{figure}
  \subfloat[max level $6$]{
    \includegraphics[width=0.45\textwidth]{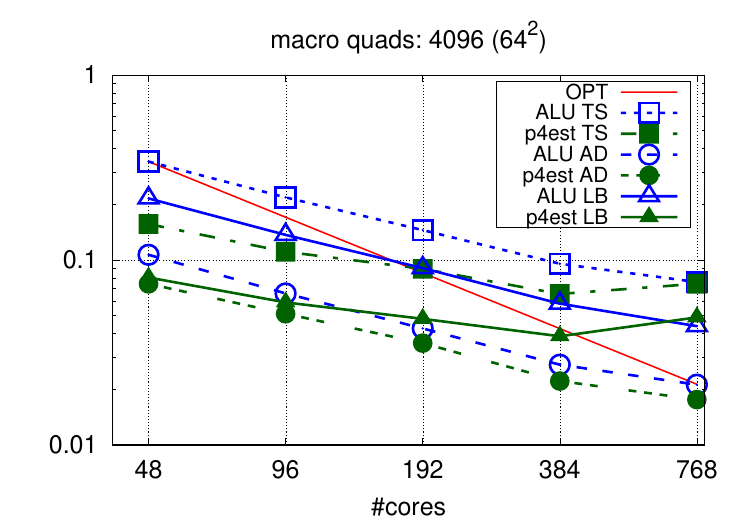}}
  \subfloat[max level $8$]{
    \includegraphics[width=0.45\textwidth]{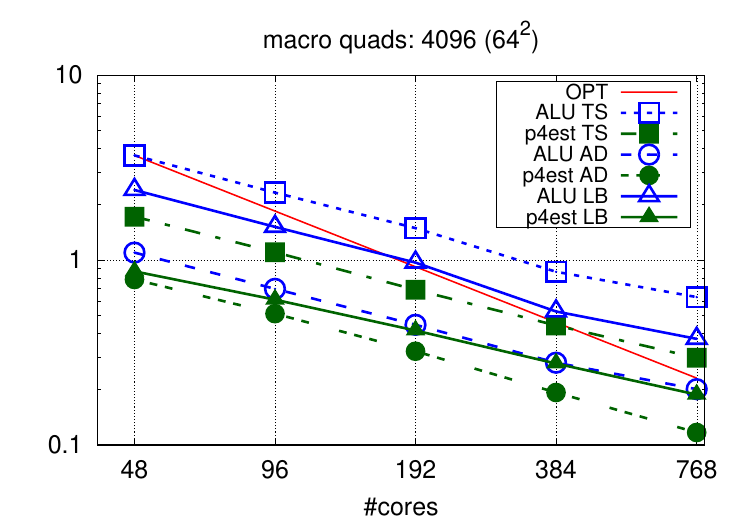}}
  \caption{Scaling results for rotating doughnut experiment in 2D for different
  maximal refinement levels. Run times for \dunealugrid are indicated by tag ALU
  and runs done with \dunepfestgrid with tag \pforest.}
  \label{fig:ball_alup4est}
\end{figure}

In Figure \ref{fig:ball_pics} snapshots of the rotating doughnut experiment for
\dunealugrid and \dunepfestgrid are presented. This experiments only considers
grid modification, no solving of PDEs is involved making this a very challenging
experiment for the grid structures. In Figure \ref{fig:ball_alup4est} we present
the strong scaling of \dunealugrid and \dunepfestgrid for a macro grid
consisting of $64^2 = 4096$ quadrilaterals (trees) and two different maximal
refinement levels. We can clearly see that \dunepfestgrid outperforms
\dunealugrid in general and in particular for the load balancing operation.
This is not surprising since the 2D implementation of \dunealugrid is based on a
specialized 3D grid as explained in \cite{alugrid:16}.

\begin{figure}
  \subfloat[max level $6$]{
    \includegraphics[width=0.45\textwidth]{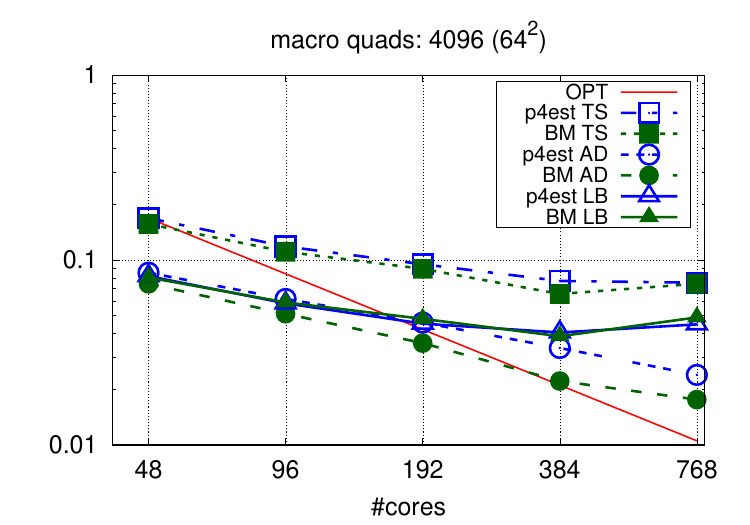}}
  \subfloat[max level $8$]{
    \includegraphics[width=0.45\textwidth]{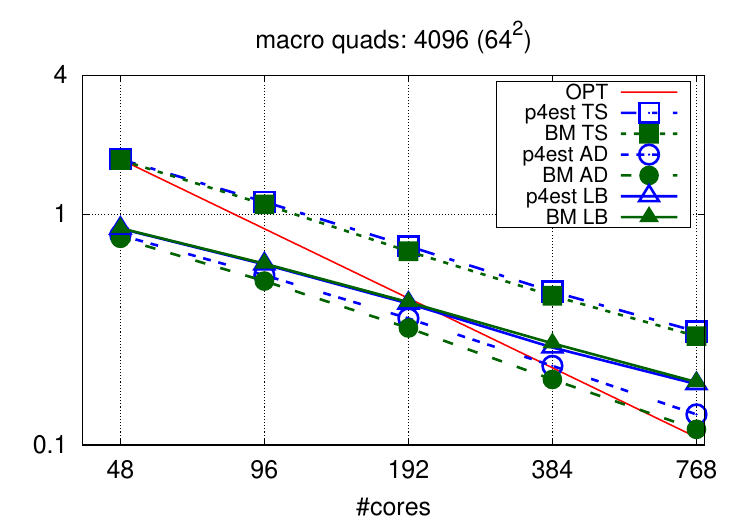}}
  \caption{Scaling results for rotating doughnut experiment in 2D for different
  maximal refinement levels. Run times for \dunepfestgrid with
  \code{p4est_balance} tagged with \pforest and with balanced marking strategy
  presented in Algorithm \ref{alg:balancedmarking} tagged BM are presented.}
  \label{fig:ball_p4est_bm}
\end{figure}

In Figure \ref{fig:ball_p4est_bm} we compare the use of \code{p4est_balance} with
the use of the balanced marking strategy described in Algorithm \ref{alg:balancedmarking}
for two different refinement levels. In both cases we can see that the balanced
marking strategy is faster than the \code{p4est_balance} which improved the
overall run time. In the experiments considered the balanced marking strategy
did not need more than $3$ iterations making it faster than the \code{p4est_balance}
application which may not be the case in all applications. This makes the
\code{p4est_balance} a stable fall back for this improved marking strategy.




\subsection{Explicit finite volume scheme}
\label{sec:finitevolume}

\renewcommand{\picturepath}{results/euler}

\begin{figure}
  \begin{center}
  \subfloat[$t=0$]{\includegraphics[width=0.3\textwidth]{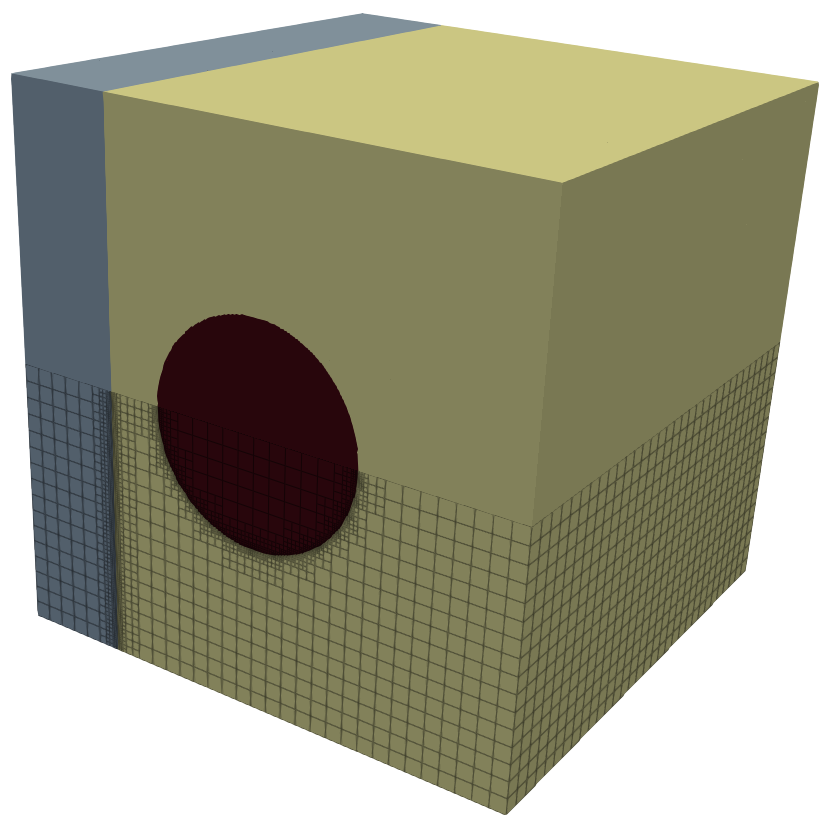}} \hfill
  \subfloat[$t=0.12$]{\includegraphics[width=0.3\textwidth]{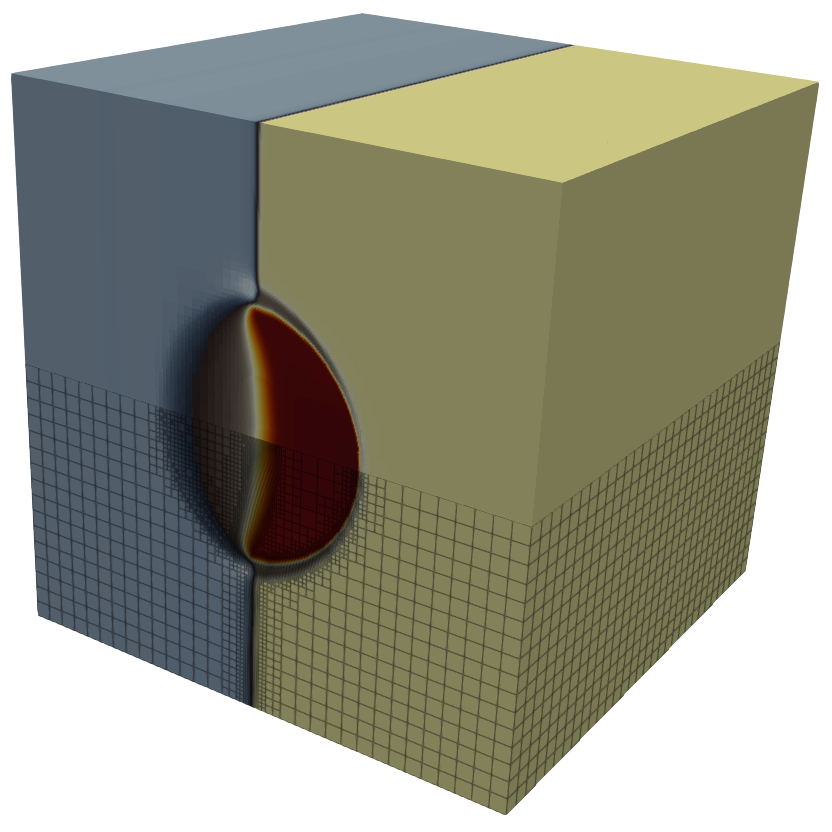}} \hfill
  \subfloat[$t=0.25$]{\includegraphics[width=0.3\textwidth]{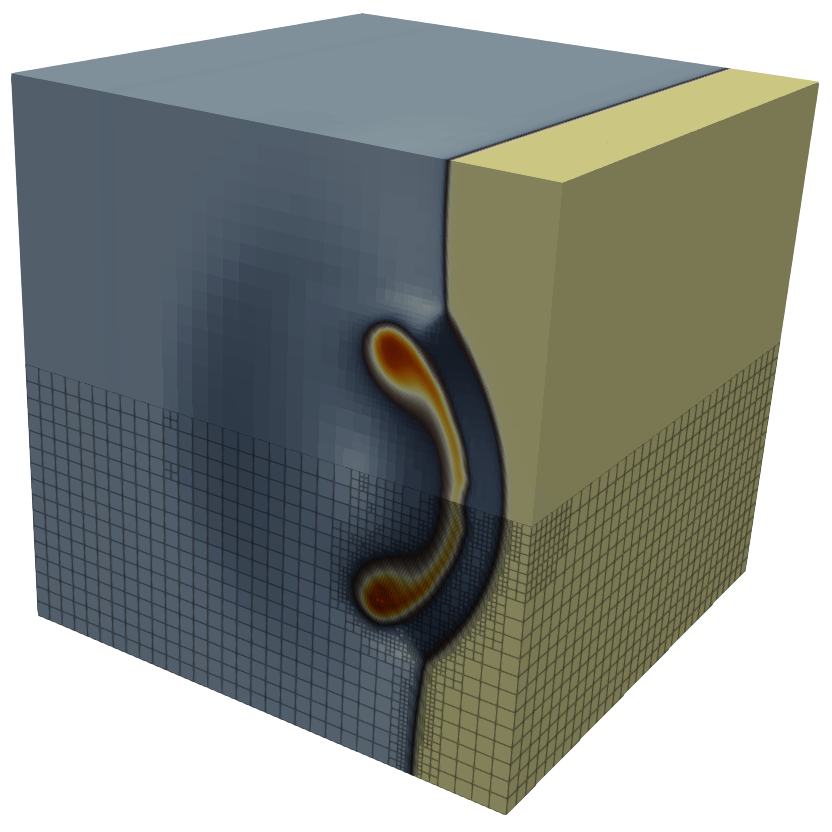}}\\
  \caption{Density of the shock bubble test case for different simulation times.
    The computation was done on $192$ cores using \dunepfestgrid and a maximal refinement level $6$.}
  \label{fig:sbresults}
  \end{center}
\end{figure}

In Figure \ref{fig:sbresults} the results of the shock bubble experiment using
\dunepfestgrid on $192$ cores with maximal refinement level of $6$ are
presented. The experiment requires frequent refinement and coarsening of grid
elements throughout the simulation making it again a very challenging experiment
for any adaptive grid manager.

\begin{figure}
  \begin{center}
  \subfloat[config $226$]{\includegraphics[width=0.48\textwidth]{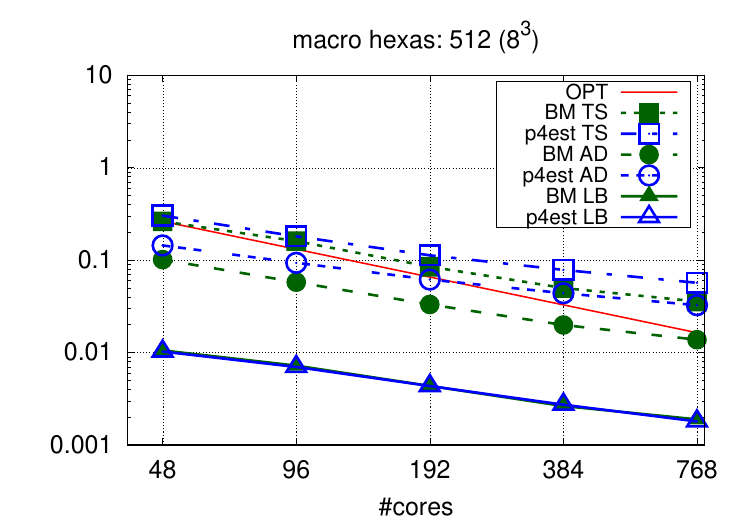}} \hfill
  \subfloat[config $204$]{\includegraphics[width=0.48\textwidth]{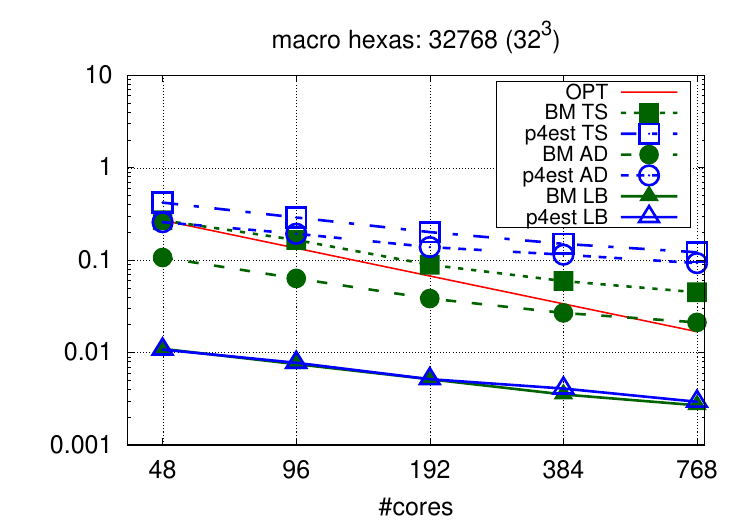}}
    \caption{Scaling for Euler shock bubble computed on cosmos with $48$ to $768$ cores. The scaling curves show computation of
    the test case with the balanced marking strategy presented in Algorithm \ref{alg:balancedmarking} (indicated by BM)
    compared to runs with the \code{p4est_balance} (indicated by \pforest).}
    \label{fig:balancedmarked1}
  \end{center}
\end{figure}

In Figure \ref{fig:balancedmarked1} we compare again the use of
\code{p4est_balance} with the strategy presented in Algorithm
\ref{alg:balancedmarking}. We can clearly see an improvement of the overall
runtime as well as the adaptation step. The load balancing (partition) step remains the same
since it is not at all affected by the change of marking strategies.

\begin{figure}
  \begin{center}
    \subfloat[config $206$]{\label{fig:balancedmarked2_a}
    \includegraphics[width=0.48\textwidth]{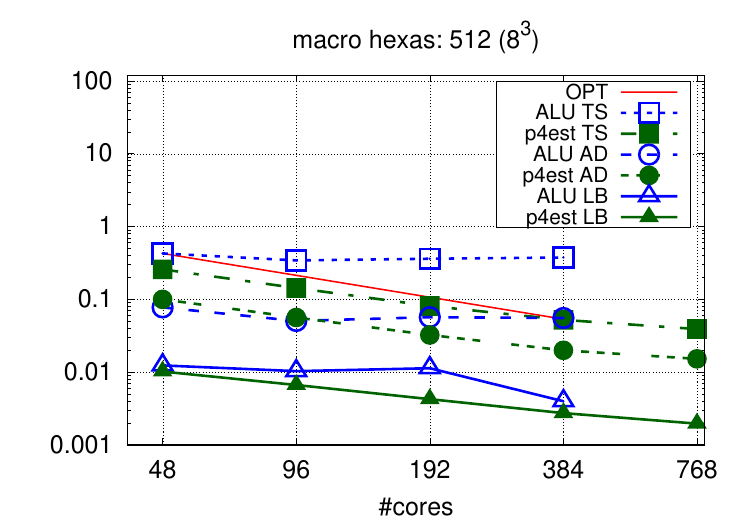}} \hfill
    \subfloat[config $204$]{\label{fig:balancedmarked2_b}
    \includegraphics[width=0.48\textwidth]{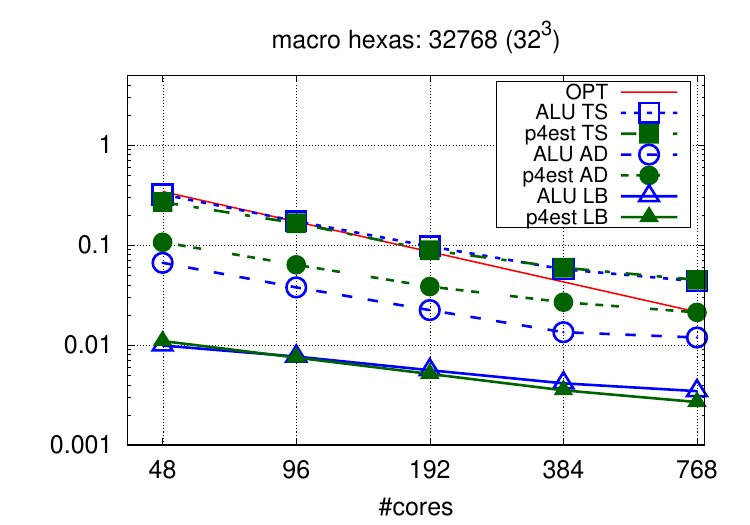}}
    \caption{Scaling for Euler shock bubble computed on COSMOS with $48$ to $768$ cores. The scaling curves show computation of
    the test case with \dunealugrid (indicated by ALU)  compared to runs with the \dunepfestgrid (indicated by \pforest).}
    \label{fig:balancedmarked2}
  \end{center}
\end{figure}

In Figure \ref{fig:balancedmarked2} we compare the run times for the shock
bubble experiment using \dunealugrid and \dunepfestgrid. When only few macro
elements are present like in Figure \ref{fig:balancedmarked2_a} we observe that
\dunealugrid scales very poorly since the partition is based on the macro grid
only. In Figure \ref{fig:balancedmarked2_b} we see that this improves when more
macro elements are present and then actually the run times between \dunealugrid
and \dunepfestgrid are quite similar. Nevertheless, \dunepfestgrid offers
several improvements, such as improved load balancing and significantly lower
memory footprint.


%% file: perfmeassures.tex
\section{Performance evaluation}
\label{sec:performancetesting}

\newcommand{\tbl}[2]{\begin{center}\caption{#1}{#2}\end{center}}
\newcommand{\setR}{\mathbb{R}}
\newcommand{\setN}{\mathbb{N}}
\newcommand{\identity}{\mathbb{I}}
\newcommand{\vecstyle}[1]{{\boldsymbol{#1}}}
\newcommand{\Grid}{\mathcal{G}}
\newcommand{\vecugrid}{\vecu_{\grid}}
\newcommand{\vecuelem}{\vecu_{\elem}}
\newcommand{\vecf}{\vecstyle{f}}
\newcommand{\vecG}{\vecstyle{G}}
\newcommand{\vecnu}{\vecstyle{\nu}}
\newcommand{\fall}{\forall \,}

We parallel the evaluation of the \dunealugrid module \cite{alugrid:16} and
use an adaptive explicit finite volume (FV) scheme for testing and comparing
the efficiency of the new \dunepfestgrid implementation against it.
FV schemes are
widely used for solving hyperbolic conservation laws. The appearance of steep
gradients or shocks in the solution make grid adaptivity a mandatory feature
for state-of-the-art
schemes. These shocks move in time requiring the
refinement zones to move with the shocks and coarsening to take place
behind them. In combination with a domain decomposition approach for
parallel computation, this means that the load is difficult to balance
a-priori between processors and dynamic load balancing is essential.
So in each time
step the grid needs to be locally refined or coarsened and the grid has to be
repartitioned quite often. What makes this problem extremely challenging is
the fact that evolving the solution from one time step to the next is very
cheap since the update is explicit and no expensive linear systems have
to be solved. So adaptivity and load balancing will dominate the
computational cost of the solver. Both of these steps require small-data
global communication as well as point-to-point messages transporting a
significant amount of data and are therefore difficult to implement at
scale.
Therefore, grid performance plays a crucial role for this problem, as it
does in any matrix-free method where frequent grid iteration occurs in order
to evaluate differential operators even if the discrete function space used
is of higher order.

In contrast, the performance of implicit matrix-based methods will have a
stronger dependency on the efficiency of the parallel solver package
than on the grid implementation.
Therefore, testing implicit methods would not provide as much insight into the
performance of the grid module itself.
For these reasons we have decided to continue using explicit finite volume
schemes as a demanding problem for a parallel grid manager to measure the
performance of the \dunepfestgrid compared to the \dune[ALUGrid] module.

As a benchmark example
we consider the Euler equations
of gas dynamics
\begin{equation*}
  \partial_t \begin{pmatrix} \rho \\ \rho\,\vec{v} \\ \epsilon \end{pmatrix}
    + \nabla \cdot \begin{pmatrix} \rho\,\vec{v} \\ \rho\,\vec{v} \otimes \vec{v} + p\,\identity \\ (\epsilon + p)\,\vec{v} \end{pmatrix}
    = 0,
\end{equation*}
where $\identity \in \setR^{d \times d}$ denotes the identity matrix.
We consider an ideal gas, i.e.,
$p = (\gamma - 1)\,(\epsilon - \frac{1}{2}\,\rho\,\lvert \vec{v} \rvert^2)$,
with the adiabatic constant $\gamma = 1.4$.
In the adaptive scheme, we use an HLLC numerical flux \citep{toro:09} in the
evolution step and the relative jump in the density to drive the grid adaptation.
Two typical test problems found in the literature, the Forward Facing Step and
the interaction between a shock and a bubble (see \citep{limiter:11} and
references therein) are implemented (see \file{examples/problem-euler.hh}).

To benchmark solely adaptation and load balancing, we implemented a third, even
more demanding test case.
Instead of using the solution to a partial differential equation to determine
the zones for grid refinement and coarsening, a simple boolean function
$\elem \mapsto \eta_{\elem}$ is used (see \file{examples/problem-ball.hh}).
We refine all elements located near the surface of a ball rotating around the
center of the 3D unit cube:
\begin{equation}
  \label{test:ball}
  \begin{aligned}
    \boldsymbol{y}( t ) &:= \Bigl( \tfrac{1}{2} + \tfrac{1}{3} \cos(2\pi t), \tfrac{1}{2} + \tfrac{1}{3} \sin(2\pi t), \tfrac{1}{2} \Bigr)^T, \\
    \eta_{\elem} &:=
    \begin{cases}
      1  &  \text{if } 0.15 < |\ics_{\elem} - \boldsymbol{y}( t )| < 0.25, \\
      0  &  \text{otherwise,}
    \end{cases}
  \end{aligned}
\end{equation}
where $\ics_{\elem}$ denotes the barycenter of the element $\elem$.
A cell $E$ is marked for refinement if $\eta_{\elem} = 1$ and for coarsening otherwise.
This sort of problem was also studied in \cite{kloefkorn:phd}.
Since the center of the ball is rotating, frequent refinement and coarsening occurs,
making this an excellent test for the implemented adaptation and load balancing
strategies.

\subsection{Definition of performance measures}
\label{perfmeasures}


For the numerical experiments in this paper two different parallel computers
were used.
\begin{description}
  \item[COSMOS\footnote{\url{https://www.lunarc.lu.se/systems/cosmos/}}]
    is a medium sized cluster
    at Lund University consisting of $182$ compute nodes with two AMD 7413
    processors (Milan) providing $2 \times 24$ cores per node.
    To compile the code the \code{foss/2024a} open source software stack was
    used.  At \textsc{COSMOS} jobs with up to $1000$ cores can be submitted.
  \item[Marvin] has recently been installed at Bonn university.
    It consists of several partitions.  For the experiments shown here,
    we use the IntelSR partition containing 192 physacil nodes with
    2 Intel Xeon "Sapphire Rapids" 48-core CPUs each.
    The theoretical maximum MPI job size on this partition is $18432$.
    The software stack is the default with OpenMPI 5.0.3.
\end{description}

\newcommand{\direct}{\pfest}
\newcommand{\bruch}{\par}

\newcommand{\deltaT}{\triangle t}
\newcommand{\deltaTn}{\deltaT^n}
\newcommand{\numleafelem}{|\grid^{n}|}
\newcommand{\N}{{\mathbbm N}}

In order to compare computations using different grids, \dunealugrid and
\dunepfestgrid, we define
performance measures for each sub-step of the numerical algorithm as follows.

For each computation (which is done on the leaf grid $\grid$
of the considered hierarchical grid $\hgrid$)
and for all computed time steps $n=0,\ldots,N-1$
the following values are stored:

\begin{center}
\begin{tabular}{lcl}
$t^n$       & -- & computational time at step $n$,      \\
$\deltaTn$  & -- & time step size for time step $n$,          \\
$\numleafelem$ & -- & number of overall interior elements of the grid at time step $n$, \\
  $\tau^n_{CO}$  & -- & time needed for communication during each time step (COMM), \\
  $\tau^n_{SO}$  & -- & time needed to compute one timestep (SOLVE), \\
  $\tau^n_{AD}$  & -- & time needed to carry out adaptation (ADAPT), \\
  $\tau^n_{LB}$  & -- & time needed to carry out load balancing (LB), and \\
  $\tau^n_{TS}$  & -- & overall time needed for time step $n$ (TS). \\
\end{tabular} \\
\end{center}

In order to compare both grid implementations
we calculate an average run time value for each computation
which is independent of the number of time steps and leaf elements.
Therefore, we compute the sum over all time steps
of the computation times needed for a sub-step
divided by the number of leaf elements of this time step, e.g.
\begin{eqnarray}
\eta_{i,\elem} := \frac{1}{N} \sum^{N-1}_{n=0}
\frac{\tau^n_i}{\numleafelem}.
\label{eq:etaruntimes}
\end{eqnarray}
with $i \in \{CO,SO,AD,LB,TS\}$.
The values of $\eta_{i,\elem}$ are then used to compute \emph{speedup} as
\begin{equation}
s_{L\to K} := \frac{\eta^L_{i,\elem}}{\eta^K_{i,\elem}},
\end{equation}
with
\begin{equation}
\eta^L_{i,\elem} := \frac{1}{L} \sum^{L-1}_{p=0} \eta^p_{i,\elem}.
\end{equation}
as the average run time of all $L$ processes.
The speedup of the parallel program
is optimal in the case $s_{L\to K} = \frac{K}{L}$.
The \emph{efficiency} from $L$ to $K > L$ processors is defined as
\begin{equation}
e_{L\to K}:= \frac{L}{K}s_{L\to K}.
\end{equation}
The efficiency of the program is called optimal if $e_{L\to K} = 1$.

For readability the scaling plots provided in Section \ref{sec:rotatingball}
and Section \ref{sec:finitevolume} show abbreviations, for
example, $TS = \eta_{TS,\elem}$ as defined in \eqref{eq:etaruntimes}, and in the
same way for $AD$ and $LB$.

%% file: conclusions.tex
\section{Conclusion}
\label{sec:conclusions}



In this submission, we present a new \dune grid implementation that can
serve as a drop-in replacement for existing implementations of \dune grids
allowing for large scale parallel AMR.
The implementation of the \dune grid interface follows closely the
\dunealugrid module and improves upon some of its shortcomings, for example,
by offering a higher scalability and lower memory footprint.  This is
clearly visible, for example, from studying the run times in 2D and an
overall improved scalability, in particular when the number of coarse grid
elements is close to the number of processes used for the simulation.

The examples shown in this article can be considered worst case AMR
scenarios where either no PDE is solved at all or a very cheap first order
finite volume algorithm is used to investigate the grid's performance.
These examples also lead to the consideration of an alternative marking
strategy that ensures a 2:1 balance between neighboring elements while
significantly reducing the run time of the program.
The authors expect to round up the results for the final version of this
document using the Marvin cluster at the University of Bonn, as well as the
shared-memory extension, which did not quite make it due to time
constraints.

As a next step and subject of subsequent publications, we will employ
\dunepfestgrid in more serious computations such as solving the
Navier-Stokes and Cahn-Hilliard equations, or other interesting examples
that require both parallel AMR and heavy computation.
We will also demonstrate that non-rectangular forest meshes work out of the
box, and extend the displayed range of coarse mesh geometries in general.
Finally, we will demonstrate numerical approaches relying on full
codimension connectivity such as generic finite element discretizations.

In this submission, we provide performance results up to 768 parallel MPI
processes.  Judging by abundant evidence, we expect the \pforest-\dune
combination to scale up for orders of magnitude more.
We are hoping to provide results for up to 18432 cores for the revised
version of this paper, depending on the state of the queuing system of a
particular supercomputer at hand.

\section*{Acknowledgements}

The authors gratefully acknowledge
access to the COSMOS compute cluster hosted at Lund University, Sweden.
Robert Kl\"ofkorn acknowledges support from the Swedish Research Council
though grant AI-Twin (2024-04904).
Mikhail Kirilin acknowledges funding by the Deutsche Forschungsgemeinschaft
(DFG, German Research Foundation) under Germany's Excellence Strategy -- The
Berlin Mathematics Research Center MATH+ (EXC-2046/1, project ID:
390685689, AA5-11).

%% file: installation.tex
\section{Installation}
\seclab{installation}

Currently, the code can only be installed directly from the source code.

The code repository is found at \url{https://gitlab.dune-project.org/extensions/dune-p4estgrid}.

A subfolder \texttt{scripts} provides convenience scripts to
build \pforest (\url{https://gitlab.dune-project.org/extensions/dune-p4estgrid/-/blob/master/scripts/build-p4est.sh?ref_type=heads}, accessed \today)
and \dunepfestgrid (\url{https://gitlab.dune-project.org/extensions/dune-p4estgrid/-/blob/master/scripts/build-dune-p4estgrid.sh?ref_type=heads}, accessed \today).

The latest master branch of \pforest (\url{https://github.com/cburstedde/p4est}, accessed \today) has to
be used with the current version of \dunepfestgrid.

\section{Examples}

Both \dunealugrid and \dunepfestgrid provide and subfolder \code{examples} which
contain the implementation of the examples used in this paper.

For \dunealugrid the particular examples used are in subfolder
\code{examples/callback} and the targets are \code{main_ball_cb_2d} or \code{main_ball_cb_3d} and
\code{main_euler_cb_2d} or \code{main_euler_cb_3d}.

The example is then run with
\begin{python}
./main_ball_cb_2d p c f none
\end{python}
where \code{p} specifies the problem number, which is $2$ is both cases and
\code{c} is the coarsest level and \code{f} the finest level allowed.
In particular, $c\in\{0,f\}$ and $f \geq c$. If \code{none} is provided then no
VTK output is produced, otherwise this parameter can be used to specify an
output folder.

For \dunepfestgrid the examples are located in the subfolder examples and target
names are \code{main_p4est_ball_2d} or \code{main_p4est_ball_3d} and
\code{main_p4est_euler_2d} or \code{main_p4est_euler_3d}.

For the experiment in Section \ref{sec:finitevolume} we used, for example,
\begin{python}
./main_p4est_euler_3d 2 0 4 none
\end{python}

%% file: paper.bbl

\begin{thebibliography}{30}
\ifx \bisbn   \undefined \def \bisbn  #1{ISBN #1}\fi
\ifx \binits  \undefined \def \binits#1{#1}\fi
\ifx \bauthor  \undefined \def \bauthor#1{#1}\fi
\ifx \batitle  \undefined \def \batitle#1{#1}\fi
\ifx \bjtitle  \undefined \def \bjtitle#1{#1}\fi
\ifx \bvolume  \undefined \def \bvolume#1{\textbf{#1}}\fi
\ifx \byear  \undefined \def \byear#1{#1}\fi
\ifx \bissue  \undefined \def \bissue#1{#1}\fi
\ifx \bfpage  \undefined \def \bfpage#1{#1}\fi
\ifx \blpage  \undefined \def \blpage #1{#1}\fi
\ifx \burl  \undefined \def \burl#1{\textsf{#1}}\fi
\ifx \doiurl  \undefined \def \doiurl#1{\url{https://doi.org/#1}}\fi
\ifx \betal  \undefined \def \betal{\textit{et al.}}\fi
\ifx \binstitute  \undefined \def \binstitute#1{#1}\fi
\ifx \binstitutionaled  \undefined \def \binstitutionaled#1{#1}\fi
\ifx \bctitle  \undefined \def \bctitle#1{#1}\fi
\ifx \beditor  \undefined \def \beditor#1{#1}\fi
\ifx \bpublisher  \undefined \def \bpublisher#1{#1}\fi
\ifx \bbtitle  \undefined \def \bbtitle#1{#1}\fi
\ifx \bedition  \undefined \def \bedition#1{#1}\fi
\ifx \bseriesno  \undefined \def \bseriesno#1{#1}\fi
\ifx \blocation  \undefined \def \blocation#1{#1}\fi
\ifx \bsertitle  \undefined \def \bsertitle#1{#1}\fi
\ifx \bsnm \undefined \def \bsnm#1{#1}\fi
\ifx \bsuffix \undefined \def \bsuffix#1{#1}\fi
\ifx \bparticle \undefined \def \bparticle#1{#1}\fi
\ifx \barticle \undefined \def \barticle#1{#1}\fi
\bibcommenthead
\ifx \bconfdate \undefined \def \bconfdate #1{#1}\fi
\ifx \botherref \undefined \def \botherref #1{#1}\fi
\ifx \url \undefined \def \url#1{\textsf{#1}}\fi
\ifx \bchapter \undefined \def \bchapter#1{#1}\fi
\ifx \bbook \undefined \def \bbook#1{#1}\fi
\ifx \bcomment \undefined \def \bcomment#1{#1}\fi
\ifx \oauthor \undefined \def \oauthor#1{#1}\fi
\ifx \citeauthoryear \undefined \def \citeauthoryear#1{#1}\fi
\ifx \endbibitem  \undefined \def \endbibitem {}\fi
\ifx \bconflocation  \undefined \def \bconflocation#1{#1}\fi
\ifx \arxivurl  \undefined \def \arxivurl#1{\textsf{#1}}\fi
\csname PreBibitemsHook\endcsname

\bibitem[\protect\citeauthoryear{Adams et~al.}{2025}]{Chombo}
\begin{botherref}
\oauthor{\bsnm{Adams}, \binits{M.}},
\oauthor{\bsnm{Colella}, \binits{P.}},
\oauthor{\bsnm{Graves}, \binits{D.T.}},
\oauthor{\bsnm{Johnson}, \binits{J.N.}},
\oauthor{\bsnm{Keen}, \binits{N.D.}},
\oauthor{\bsnm{Ligocki}, \binits{T.J.}},
\oauthor{\bsnm{Martin}, \binits{D.F.}},
\oauthor{\bsnm{McCorquodale}, \binits{P.W.}},
\oauthor{\bsnm{Modiano}, \binits{D.}},
\oauthor{\bsnm{Schwartz}, \binits{P.O.}},
\oauthor{\bsnm{Sternberg}, \binits{T.D.}},
\oauthor{\bsnm{Straalen}, \binits{B.V.}}:
Chombo software package for {AMR} applications -- design document.
Technical Report LBNL-6616E,
Lawrence Berkeley National Laboratory
(2025).
\doiurl{10.6084/m9.figshare.28599755.v1}
\end{botherref}
\endbibitem

\bibitem[\protect\citeauthoryear{Zhang et~al.}{2019}]{AMReX_JOSS}
\begin{barticle}
\bauthor{\bsnm{Zhang}, \binits{W.}},
\bauthor{\bsnm{Almgren}, \binits{A.}},
\bauthor{\bsnm{Beckner}, \binits{V.}},
\bauthor{\bsnm{Bell}, \binits{J.}},
\bauthor{\bsnm{Blaschke}, \binits{J.}},
\bauthor{\bsnm{Chan}, \binits{C.}},
\bauthor{\bsnm{Day}, \binits{M.}},
\bauthor{\bsnm{Friesen}, \binits{B.}},
\bauthor{\bsnm{Gott}, \binits{K.}},
\bauthor{\bsnm{Graves}, \binits{D.}},
\bauthor{\bsnm{Katz}, \binits{M.}},
\bauthor{\bsnm{Myers}, \binits{A.}},
\bauthor{\bsnm{Nguyen}, \binits{T.}},
\bauthor{\bsnm{Nonaka}, \binits{A.}},
\bauthor{\bsnm{Rosso}, \binits{M.}},
\bauthor{\bsnm{Williams}, \binits{S.}},
\bauthor{\bsnm{Zingale}, \binits{M.}}:
\batitle{{AMReX}: A framework for block-structured adaptive mesh refinement}.
\bjtitle{Journal of Open Source Software}
\bvolume{4}(\bissue{37}),
\bfpage{1370}
(\byear{2019})
\doiurl{10.21105/joss.01370}
\end{barticle}
\endbibitem

\bibitem[\protect\citeauthoryear{Balay et~al.}{2021}]{petsc-user-ref}
\begin{botherref}
\oauthor{\bsnm{Balay}, \binits{S.}},
\oauthor{\bsnm{Abhyankar}, \binits{S.}},
\oauthor{\bsnm{Adams}, \binits{M.F.}},
\oauthor{\bsnm{Brown}, \binits{J.}},
\oauthor{\bsnm{Brune}, \binits{P.}},
\oauthor{\bsnm{Buschelman}, \binits{K.}},
\oauthor{\bsnm{Dalcin}, \binits{L.}},
\oauthor{\bsnm{Dener}, \binits{A.}},
\oauthor{\bsnm{Eijkhout}, \binits{V.}},
\oauthor{\bsnm{Gropp}, \binits{W.D.}},
\oauthor{\bsnm{Karpeyev}, \binits{D.}},
\oauthor{\bsnm{Kaushik}, \binits{D.}},
\oauthor{\bsnm{Knepley}, \binits{M.G.}},
\oauthor{\bsnm{May}, \binits{D.A.}},
\oauthor{\bsnm{McInnes}, \binits{L.C.}},
\oauthor{\bsnm{Mills}, \binits{R.T.}},
\oauthor{\bsnm{Munson}, \binits{T.}},
\oauthor{\bsnm{Rupp}, \binits{K.}},
\oauthor{\bsnm{Sanan}, \binits{P.}},
\oauthor{\bsnm{Smith}, \binits{B.F.}},
\oauthor{\bsnm{Zampini}, \binits{S.}},
\oauthor{\bsnm{Zhang}, \binits{H.}},
\oauthor{\bsnm{Zhang}, \binits{H.}}:
{PETS}c users manual.
Technical Report ANL-95/11 -- Revision 3.15,
Argonne National Laboratory
(2021).
\url{https://www.mcs.anl.gov/petsc}
\end{botherref}
\endbibitem

\bibitem[\protect\citeauthoryear{Bangerth
  et~al.}{2011}]{BangerthBursteddeHeisterEtAl11}
\begin{barticle}
\bauthor{\bsnm{Bangerth}, \binits{W.}},
\bauthor{\bsnm{Burstedde}, \binits{C.}},
\bauthor{\bsnm{Heister}, \binits{T.}},
\bauthor{\bsnm{Kronbichler}, \binits{M.}}:
\batitle{Algorithms and data structures for massively parallel generic adaptive
  finite element codes}.
\bjtitle{{ACM} Transactions on Mathematical Software}
\bvolume{38}(\bissue{14}),
\bfpage{1}--\blpage{28}
(\byear{2011})
\doiurl{10.1145/2049673.2049678}
\end{barticle}
\endbibitem

\bibitem[\protect\citeauthoryear{MacNeice et~al.}{2011}]{paramesh}
\begin{botherref}
\oauthor{\bsnm{MacNeice}, \binits{P.}},
\oauthor{\bsnm{Olson}, \binits{K.M.}},
\oauthor{\bsnm{Mobarry}, \binits{C.}},
\oauthor{\bsnm{Fainchtein}, \binits{R.}},
\oauthor{\bsnm{Packer}, \binits{C.}}:
{PARAMESH V4.1}: Parallel adaptive mesh refinement.
Technical Report 1106.009,
Astrophysics Source Code Library
(2011)
\end{botherref}
\endbibitem

\bibitem[\protect\citeauthoryear{Bastian et~al.}{2021}]{dunereview:21}
\begin{barticle}
\bauthor{\bsnm{Bastian}, \binits{P.}},
\bauthor{\bsnm{Blatt}, \binits{M.}},
\bauthor{\bsnm{Dedner}, \binits{M.}},
\bauthor{\bsnm{Dreier}, \binits{N.-A.}},
\bauthor{\bsnm{Engwer}, \binits{R.} \bsuffix{Ch.~Fritze}},
\bauthor{\bsnm{Gr{\"a}ser}, \binits{C.}},
\bauthor{\bsnm{Gr{\"u}ninger}, \binits{C.}},
\bauthor{\bsnm{Kempf}, \binits{D.}},
\bauthor{\bsnm{Kl{\"o}fkorn}, \binits{R.}},
\bauthor{\bsnm{Ohlberger}, \binits{M.}},
\bauthor{\bsnm{Sander}, \binits{O.}}:
\batitle{The {Dune} framework: Basic concepts and recent developments}.
\bjtitle{{CAMWA}}
\bvolume{81},
\bfpage{75}--\blpage{112}
(\byear{2021})
\doiurl{10.1016/j.camwa.2020.06.007}
\end{barticle}
\endbibitem

\bibitem[\protect\citeauthoryear{Burstedde
  et~al.}{2011}]{BursteddeWilcoxGhattas11}
\begin{barticle}
\bauthor{\bsnm{Burstedde}, \binits{C.}},
\bauthor{\bsnm{Wilcox}, \binits{L.C.}},
\bauthor{\bsnm{Ghattas}, \binits{O.}}:
\batitle{{\texttt{\upshape p4est}}: Scalable algorithms for parallel adaptive
  mesh refinement on forests of octrees}.
\bjtitle{SIAM Journal on Scientific Computing}
\bvolume{33}(\bissue{3}),
\bfpage{1103}--\blpage{1133}
(\byear{2011})
\doiurl{10.1137/100791634}
\end{barticle}
\endbibitem

\bibitem[\protect\citeauthoryear{Rudi et~al.}{2015}]{RudiMalossiIsaacEtAl15}
\begin{bchapter}
\bauthor{\bsnm{Rudi}, \binits{J.}},
\bauthor{\bsnm{Malossi}, \binits{A.C.I.}},
\bauthor{\bsnm{Isaac}, \binits{T.}},
\bauthor{\bsnm{Stadler}, \binits{G.}},
\bauthor{\bsnm{Gurnis}, \binits{M.}},
\bauthor{\bsnm{Staar}, \binits{P.W.J.}},
\bauthor{\bsnm{Ineichen}, \binits{Y.}},
\bauthor{\bsnm{Bekas}, \binits{C.}},
\bauthor{\bsnm{Curioni}, \binits{A.}},
\bauthor{\bsnm{Ghattas}, \binits{O.}}:
\bctitle{An extreme-scale implicit solver for complex {PDE}s: Highly
  heterogeneous flow in earth's mantle}.
In: \bbtitle{Proceedings of the International Conference for High Performance
  Computing, Networking, Storage and Analysis}.
\bsertitle{SC '15}.
\bpublisher{Association for Computing Machinery},
\blocation{New York, NY, USA}
(\byear{2015}).
\doiurl{10.1145/2807591.2807675} .
\burl{https://doi.org/10.1145/2807591.2807675}
\end{bchapter}
\endbibitem

\bibitem[\protect\citeauthoryear{Bastian et~al.}{2008}]{dunepaperII:08}
\begin{barticle}
\bauthor{\bsnm{Bastian}, \binits{P.}},
\bauthor{\bsnm{Blatt}, \binits{M.}},
\bauthor{\bsnm{Dedner}, \binits{A.}},
\bauthor{\bsnm{Engwer}, \binits{C.}},
\bauthor{\bsnm{Kl{\"o}fkorn}, \binits{R.}},
\bauthor{\bsnm{Kornhuber}, \binits{R.}},
\bauthor{\bsnm{Ohlberger}, \binits{M.}},
\bauthor{\bsnm{Sander}, \binits{O.}}:
\batitle{A generic grid interface for parallel and adaptive scientific
  computing. part {II}: Implementation and tests in {DUNE}}.
\bjtitle{Computing}
\bvolume{82}(\bissue{2--3}),
\bfpage{121}--\blpage{138}
(\byear{2008})
\doiurl{10.1007/s00607-008-0004-9}
\end{barticle}
\endbibitem

\bibitem[\protect\citeauthoryear{Alk{\"a}mper et~al.}{2016}]{alugrid:16}
\begin{barticle}
\bauthor{\bsnm{Alk{\"a}mper}, \binits{M.}},
\bauthor{\bsnm{Dedner}, \binits{A.}},
\bauthor{\bsnm{Kl{\"o}fkorn}, \binits{R.}},
\bauthor{\bsnm{Nolte}, \binits{M.}}:
\batitle{{The DUNE-ALUGrid Module.}}
\bjtitle{Archive of Numerical Software}
\bvolume{4}(\bissue{1}),
\bfpage{1}--\blpage{28}
(\byear{2016})
\doiurl{10.11588/ans.2016.1.23252}
\end{barticle}
\endbibitem

\bibitem[\protect\citeauthoryear{Bastian et~al.}{1997}]{Bastian1997}
\begin{barticle}
\bauthor{\bsnm{Bastian}, \binits{P.}},
\bauthor{\bsnm{Birken}, \binits{K.}},
\bauthor{\bsnm{Johannsen}, \binits{K.}},
\bauthor{\bsnm{Lang}, \binits{S.}},
\bauthor{\bsnm{Neu{\ss}}, \binits{N.}},
\bauthor{\bsnm{Rentz-Reichert}, \binits{H.}},
\bauthor{\bsnm{Wieners}, \binits{C.}}:
\batitle{{UG -- A flexible software toolbox for solving partial differential
  equations}}.
\bjtitle{Computing and Visualization in Science}
\bvolume{1}(\bissue{1}),
\bfpage{27}--\blpage{40}
(\byear{1997})
\doiurl{10.1007/s007910050003}
\end{barticle}
\endbibitem

\bibitem[\protect\citeauthoryear{Morton}{1966}]{Morton66}
\begin{botherref}
\oauthor{\bsnm{Morton}, \binits{G.M.}}:
A computer oriented geodetic data base; and a new technique in file sequencing.
Technical report,
IBM Ltd.
(1966)
\end{botherref}
\endbibitem

\bibitem[\protect\citeauthoryear{Tu et~al.}{2005}]{TuOHallaronGhattas05}
\begin{bchapter}
\bauthor{\bsnm{Tu}, \binits{T.}},
\bauthor{\bsnm{O'Hallaron}, \binits{D.R.}},
\bauthor{\bsnm{Ghattas}, \binits{O.}}:
\bctitle{Scalable parallel octree meshing for terascale applications}.
In: \bbtitle{SC '05: Proceedings of the International Conference for High
  Performance Computing, Networking, Storage, and Analysis},
pp. \bfpage{4}--\blpage{4}.
\bpublisher{ACM/IEEE}
(\byear{2005}).
\doiurl{10.1109/SC.2005.61}
\end{bchapter}
\endbibitem

\bibitem[\protect\citeauthoryear{Sundar et~al.}{2008}]{SundarSampathBiros08}
\begin{barticle}
\bauthor{\bsnm{Sundar}, \binits{H.}},
\bauthor{\bsnm{Sampath}, \binits{R.}},
\bauthor{\bsnm{Biros}, \binits{G.}}:
\batitle{Bottom-up construction and 2:1 balance refinement of linear octrees in
  parallel}.
\bjtitle{SIAM Journal on Scientific Computing}
\bvolume{30}(\bissue{5}),
\bfpage{2675}--\blpage{2708}
(\byear{2008})
\doiurl{10.1137/070681727}
\end{barticle}
\endbibitem

\bibitem[\protect\citeauthoryear{Stewart and Edwards}{2004}]{StewartEdwards04}
\begin{barticle}
\bauthor{\bsnm{Stewart}, \binits{J.R.}},
\bauthor{\bsnm{Edwards}, \binits{H.C.}}:
\batitle{A framework approach for developing parallel adaptive multiphysics
  applications}.
\bjtitle{Finite Elements in Analysis and Design}
\bvolume{40}(\bissue{12}),
\bfpage{1599}--\blpage{1617}
(\byear{2004})
\doiurl{10.1016/j.finel.2003.10.006}
\end{barticle}
\endbibitem

\bibitem[\protect\citeauthoryear{Bangerth
  et~al.}{2007}]{BangerthHartmannKanschat07}
\begin{barticle}
\bauthor{\bsnm{Bangerth}, \binits{W.}},
\bauthor{\bsnm{Hartmann}, \binits{R.}},
\bauthor{\bsnm{Kanschat}, \binits{G.}}:
\batitle{deal.{II} -- a general-purpose object-oriented finite element
  library}.
\bjtitle{ACM Transactions on Mathematical Software}
\bvolume{33}(\bissue{4}),
\bfpage{24}
(\byear{2007})
\doiurl{10.1145/1268776.1268779}
\end{barticle}
\endbibitem

\bibitem[\protect\citeauthoryear{Burstedde and Holke}{2015}]{BursteddeHolke15}
\begin{botherref}
\oauthor{\bsnm{Burstedde}, \binits{C.}},
\oauthor{\bsnm{Holke}, \binits{J.}}:
A tetrahedral space-filling curve for non-conforming adaptive meshes
(2015)
\end{botherref}
\endbibitem

\bibitem[\protect\citeauthoryear{Burstedde}{2010}]{Burstedde25a}
\begin{botherref}
\oauthor{\bsnm{Burstedde}, \binits{C.}}:
{\texttt{\upshape p4est}}: Parallel {AMR} on Forests of Octrees.
\url{https://www.p4est.org/} (last accessed June 18th, 2025)
(2010)
\end{botherref}
\endbibitem

\bibitem[\protect\citeauthoryear{Holke
  et~al.}{2023}]{HolkeBursteddeKnappEtAl23}
\begin{bchapter}
\bauthor{\bsnm{Holke}, \binits{J.}},
\bauthor{\bsnm{Burstedde}, \binits{C.}},
\bauthor{\bsnm{Knapp}, \binits{D.}},
\bauthor{\bsnm{Dreyer}, \binits{L.}},
\bauthor{\bsnm{Elsweijer}, \binits{S.}},
\bauthor{\bsnm{{\"U}nl{\"u}}, \binits{V.}},
\bauthor{\bsnm{Markert}, \binits{J.}},
\bauthor{\bsnm{Lilikakis}, \binits{I.}},
\bauthor{\bsnm{B{\"o}ing}, \binits{N.}},
\bauthor{\bsnm{Ponnusamy}, \binits{P.}},
\bauthor{\bsnm{Basermann}, \binits{A.}}:
\bctitle{\texttt{\upshape t8code} v.\ 1.0 -- modular adaptive mesh refinement
  in the exascale era}.
In: \bbtitle{{SIAM} International Meshing Round Table 2023}.
\bpublisher{{SIAM}},
\blocation{Amsterdam, NL}
(\byear{2023}).
\burl{https://elib.dlr.de/194377/}
\end{bchapter}
\endbibitem

\bibitem[\protect\citeauthoryear{Burstedde and Holke}{2016}]{BursteddeHolke16b}
\begin{bchapter}
\bauthor{\bsnm{Burstedde}, \binits{C.}},
\bauthor{\bsnm{Holke}, \binits{J.}}:
\bctitle{\texttt{\upshape p4est}: Scalable algorithms for parallel adaptive
  mesh refinement}.
In: \beditor{\bsnm{Br{\"o}mmel}, \binits{D.}},
\beditor{\bsnm{Frings}, \binits{W.}},
\beditor{\bsnm{Wylie}, \binits{B.J.N.}} (eds.)
\bbtitle{JUQUEEN Extreme Scaling Workshop 2016}.
\bsertitle{JSC Internal Report},
pp. \bfpage{49}--\blpage{54}.
\bpublisher{J\"ulich Supercomputing Centre}
(\byear{2016})
\end{bchapter}
\endbibitem

\bibitem[\protect\citeauthoryear{Isaac
  et~al.}{2015}]{IsaacBursteddeWilcoxEtAl15}
\begin{barticle}
\bauthor{\bsnm{Isaac}, \binits{T.}},
\bauthor{\bsnm{Burstedde}, \binits{C.}},
\bauthor{\bsnm{Wilcox}, \binits{L.C.}},
\bauthor{\bsnm{Ghattas}, \binits{O.}}:
\batitle{Recursive algorithms for distributed forests of octrees}.
\bjtitle{SIAM Journal on Scientific Computing}
\bvolume{37}(\bissue{5}),
\bfpage{497}--\blpage{531}
(\byear{2015})
\doiurl{10.1137/140970963}
{\href{https://arxiv.org/abs/1406.0089}{{arXiv:1406.0089}}}
{[cs.DC]}
\end{barticle}
\endbibitem

\bibitem[\protect\citeauthoryear{Holke et~al.}{2021}]{HolkeKnappBurstedde21}
\begin{barticle}
\bauthor{\bsnm{Holke}, \binits{J.}},
\bauthor{\bsnm{Knapp}, \binits{D.}},
\bauthor{\bsnm{Burstedde}, \binits{C.}}:
\batitle{An optimized, parallel computation of the ghost layer for adaptive
  hybrid forest meshes}.
\bjtitle{SIAM Journal on Scientific Computing}
\bvolume{43}(\bissue{6}),
\bfpage{359}--\blpage{385}
(\byear{2021})
\doiurl{10.1137/20M1383033}
\end{barticle}
\endbibitem

\bibitem[\protect\citeauthoryear{Isaac et~al.}{2012}]{IsaacBursteddeGhattas12}
\begin{bchapter}
\bauthor{\bsnm{Isaac}, \binits{T.}},
\bauthor{\bsnm{Burstedde}, \binits{C.}},
\bauthor{\bsnm{Ghattas}, \binits{O.}}:
\bctitle{Low-cost parallel algorithms for 2:1 octree balance}.
In: \bbtitle{Proceedings of the 26th IEEE International Parallel {\&}
  Distributed Processing Symposium},
vol. \bseriesno{1},
pp. \bfpage{426}--\blpage{437}.
\bpublisher{IEEE}
(\byear{2012}).
\doiurl{10.1109/IPDPS.2012.47}
\end{bchapter}
\endbibitem

\bibitem[\protect\citeauthoryear{Suh and Isaac}{2020}]{SuhIsaac20}
\begin{bchapter}
\bauthor{\bsnm{Suh}, \binits{H.}},
\bauthor{\bsnm{Isaac}, \binits{T.}}:
\bctitle{Evaluation of a minimally synchronous algorithm for 2:1 octree
  balance}.
In: \bbtitle{Proceedings of the International Conference for High Performance
  Computing, Networking, Storage and Analysis}.
\bsertitle{SC {'20}}.
\bpublisher{IEEE Press}
(\byear{2020})
\end{bchapter}
\endbibitem

\bibitem[\protect\citeauthoryear{Mirzadeh
  et~al.}{2016}]{MirzadehGuittetBursteddeEtAl16}
\begin{barticle}
\bauthor{\bsnm{Mirzadeh}, \binits{M.}},
\bauthor{\bsnm{Guittet}, \binits{A.}},
\bauthor{\bsnm{Burstedde}, \binits{C.}},
\bauthor{\bsnm{Gibou}, \binits{F.}}:
\batitle{Parallel level-set methods on adaptive tree-based grids}.
\bjtitle{Journal of Computational Physics}
\bvolume{322},
\bfpage{345}--\blpage{364}
(\byear{2016})
\end{barticle}
\endbibitem

\bibitem[\protect\citeauthoryear{Kl{\"o}fkorn}{2009}]{kloefkorn:phd}
\begin{bbook}
\bauthor{\bsnm{Kl{\"o}fkorn}, \binits{R.}}:
\bbtitle{Numerics for Evolution Equations --- A General Interface Based Design
  Concept}.
\bpublisher{PhD thesis, Albert-Ludwigs-Universit{\"a}t Freiburg}
(\byear{2009}).
\doiurl{10.6094/UNIFR/7175}
\end{bbook}
\endbibitem

\bibitem[\protect\citeauthoryear{Burstedde
  et~al.}{2025a}]{BursteddeGriesbachBrandtEtAl25}
\begin{botherref}
\oauthor{\bsnm{Burstedde}, \binits{C.}},
\oauthor{\bsnm{Griesbach}, \binits{T.}},
\oauthor{\bsnm{Brandt}, \binits{H.}},
\oauthor{\bsnm{Hirsch}, \binits{M.}},
\oauthor{\bsnm{Kestener}, \binits{P.}},
\oauthor{\bsnm{Kirilin}, \binits{M.}},
\oauthor{\bsnm{Dutka}, \binits{A.}},
\oauthor{\bsnm{Elsweijer}, \binits{S.}},
\oauthor{\bsnm{Knapp}, \binits{D.}},
\oauthor{\bsnm{Markert}, \binits{J.}}:
The {\texttt{\upshape p4est}} software library.
Zenodo.
Software release version 2.8.7, \url{https://doi.org/10.5281/zenodo.15048174}
(2025).
\doiurl{10.5281/zenodo.15048174} .
\url{https://doi.org/10.5281/zenodo.15048174}
\end{botherref}
\endbibitem

\bibitem[\protect\citeauthoryear{Burstedde
  et~al.}{2025b}]{BursteddeGriesbachHirschEtAl25}
\begin{botherref}
\oauthor{\bsnm{Burstedde}, \binits{C.}},
\oauthor{\bsnm{Griesbach}, \binits{T.}},
\oauthor{\bsnm{Hirsch}, \binits{M.}},
\oauthor{\bsnm{Dutka}, \binits{A.}},
\oauthor{\bsnm{Brandt}, \binits{H.}},
\oauthor{\bsnm{Knapp}, \binits{D.}},
\oauthor{\bsnm{Kirilin}, \binits{M.}},
\oauthor{\bsnm{Karembe}, \binits{M.}},
\oauthor{\bsnm{Williams}, \binits{K.}}:
The {\texttt{\upshape sc}} software library.
Zenodo.
Software release version 2.8.7, \url{https://doi.org/10.5281/zenodo.15047905}
(2025).
\doiurl{10.5281/zenodo.15047905} .
\url{https://doi.org/10.5281/zenodo.15047905}
\end{botherref}
\endbibitem

\bibitem[\protect\citeauthoryear{Toro}{2009}]{toro:09}
\begin{bbook}
\bauthor{\bsnm{Toro}, \binits{E.}}:
\bbtitle{Riemann Solvers and Numerical Methods for Fluid Dynamics}.
\bpublisher{Springer}
(\byear{2009})
\end{bbook}
\endbibitem

\bibitem[\protect\citeauthoryear{Dedner and Kl{\"o}fkorn}{2011}]{limiter:11}
\begin{barticle}
\bauthor{\bsnm{Dedner}, \binits{A.}},
\bauthor{\bsnm{Kl{\"o}fkorn}, \binits{R.}}:
\batitle{A generic stabilization approach for higher order discontinuous
  {G}alerkin methods for convection dominated problems}.
\bjtitle{J. Sci. Comput.}
\bvolume{47}(\bissue{3}),
\bfpage{365}--\blpage{388}
(\byear{2011})
\doiurl{10.1007/s10915-010-9448-0}
\end{barticle}
\endbibitem

\end{thebibliography}
